\documentclass[12pt,preprint]{aastex} 
\usepackage{flushrt}

\slugcomment{Revised version submitted to the Astrophysical Journal}

\shortauthors{Chieffi, Limongi} \shorttitle{Massive Stars: Evolution and Nucleosynthesis}

\begin{document}

\title{THE EXPLOSIVE YIELDS PRODUCED BY THE FIRST GENERATION OF CORE COLLAPSE SUPERNOVAE AND THE CHEMICAL COMPOSITION OF EXTREMELY 
METAL POOR STARS}

\author{Alessandro Chieffi\altaffilmark{1,2} and Marco Limongi\altaffilmark{2}}

\affil{1. Istituto di Astrofisica Spaziale e Fisica Cosmica (CNR), Via Fosso del Cavaliere, I-00133, Roma, Italy; 
achieffi@ias.rm.cnr.it} 

\affil{2. Istituto Nazionale di AstroFisica - Osservatorio Astronomico di Roma, Via Frascati 33, I-00040, Monteporzio Catone, Italy; 
marco@mporzio.astro.it}

\begin{abstract} We present a detailed comparison between an extended set of elemental abundances observed in some of the most metal 
poor stars presently known and the ejecta produced by a generation of primordial core collapse supernovae. At variance with most of 
the analysis performed up to now (in which mainly the global trends with the overall metallicity are discussed), we think to be 
important (and complementary to the other approach) to fit the (available) chemical composition of specific stars. In particular we 
firstly discuss the differences among the five stars which form our initial database and define a "template" ultra metal poor star 
which is then compared to the theoretical predictions. Our main findings are as follows: a) the fit to [Si/Mg] and [Ca/Mg] of these 
very metal poor stars seems to favor the presence of a rather large C abundance at the end of the central He burning; in a classical 
scenario in which the border of the convective core is strictly determined by the Schwarzschild criterion, such a large C abundance 
would imply a rather low $\rm ^{12}C(\alpha,\gamma)^{16}O$ reaction rate; b) a low C abundance left by the central He burning would 
imply a low [Al/Mg] ($<-1.2$ dex) independently on the initial mass of the exploding star while a rather large C abundance would 
produce such a low [Al/Mg] only for the most massive stellar model; c) at variance with current beliefs that it is difficult to 
interpret the observed overabundance of [Co/Fe], we find that a mildly large C abundance in the He exhausted core (well within the 
present range of uncertainty) easily and naturally allows a very good fit to [Co/Fe]; d) our yields allow a reasonable fit to 8 
out of the 11 available elemental abundances; e) within the present grid of models it is not possible to find a good match of the 
remaining three elements, Ti, Cr and Ni (even for an arbitrary choice of the mass cut); f) the adoption of other yields available in 
the literature does not improve the fit; g) since no mass in our grid provides a satisfactory fit to these three elements, even an 
arbitrary choice of the initial mass function would not improve their fit.

\end{abstract}



\section{Introduction} Extremely metal poor stars are among the oldest objects in our Galaxy and hence the analysis of their surface 
chemical composition provides us with invaluable information about the early chemical enrichment of the pristine material. Since 
these stars probably formed in an environment enriched by just the first generation of stars, they also constitute a unique 
opportunity of observing directly the ejecta of a single stellar generation and not the complex superimposition of many generations 
of stars of different metallicity (as it happens when looking at stars of higher metallicity). Moreover, recent sets of 
observational data \citep{MPSS95,RNB96} have shown that below $\rm [Fe/H]\simeq-2.5$ it exists a significative star to star scatter 
in the elemental abundances of many stars. This scatter has been interpreted \citep{AS95} as a consequence of the fact that these 
stars formed in a highly inhomogeneous medium composed of substructures enriched by few supernovae (even one). \cite{ST98} and 
\cite{Netal99} tried to interpret the chemical composition of the most metal poor stars in terms of a "single" supernova event. 
However, both attempts did not try to fit the surface composition of any of these metal poor stars but, instead, they adopted this 
scenario to try to reproduce the average trend with the metallicity of a few elemental ratios (i.e. [Cr/Fe], [Mn/Fe],[Co/Fe]). More 
recently, \cite{asgt00} introduced a chemical evolution model able to resolve the formation of the initial chemical inhomogeneities 
and the following progressive homogenization but, also in this case, their main objective was to try to fit the observed spread of a 
few elements as a function of the iron abundance. We think it is time to test more thoroughly the idea that the most metal poor 
stars could have been formed by matter enriched by the ejecta of a single supernova event. In particular, by taking advantage of the 
fact that a rather large number of elemental abundances are available per each of these very metal poor stars, we think to be worth 
to directly compare these observed abundances to the ejecta of single primordial core collapse supernovae: the observational 
database we will analyze in this paper is formed by the five extremely metal poor stars (i.e. stars having $\rm [Fe/H]<-3.3$) 
recently published by \cite{NRB01}. The outline of the paper is as follows: the observational database will be discussed in section 
2 while the theoretical yields will be presented in section 3. Section 4 is devoted to the comparison between the observational and 
theoretical data. A final discussion and conclusion follows.

\begin{figure}
\plotone{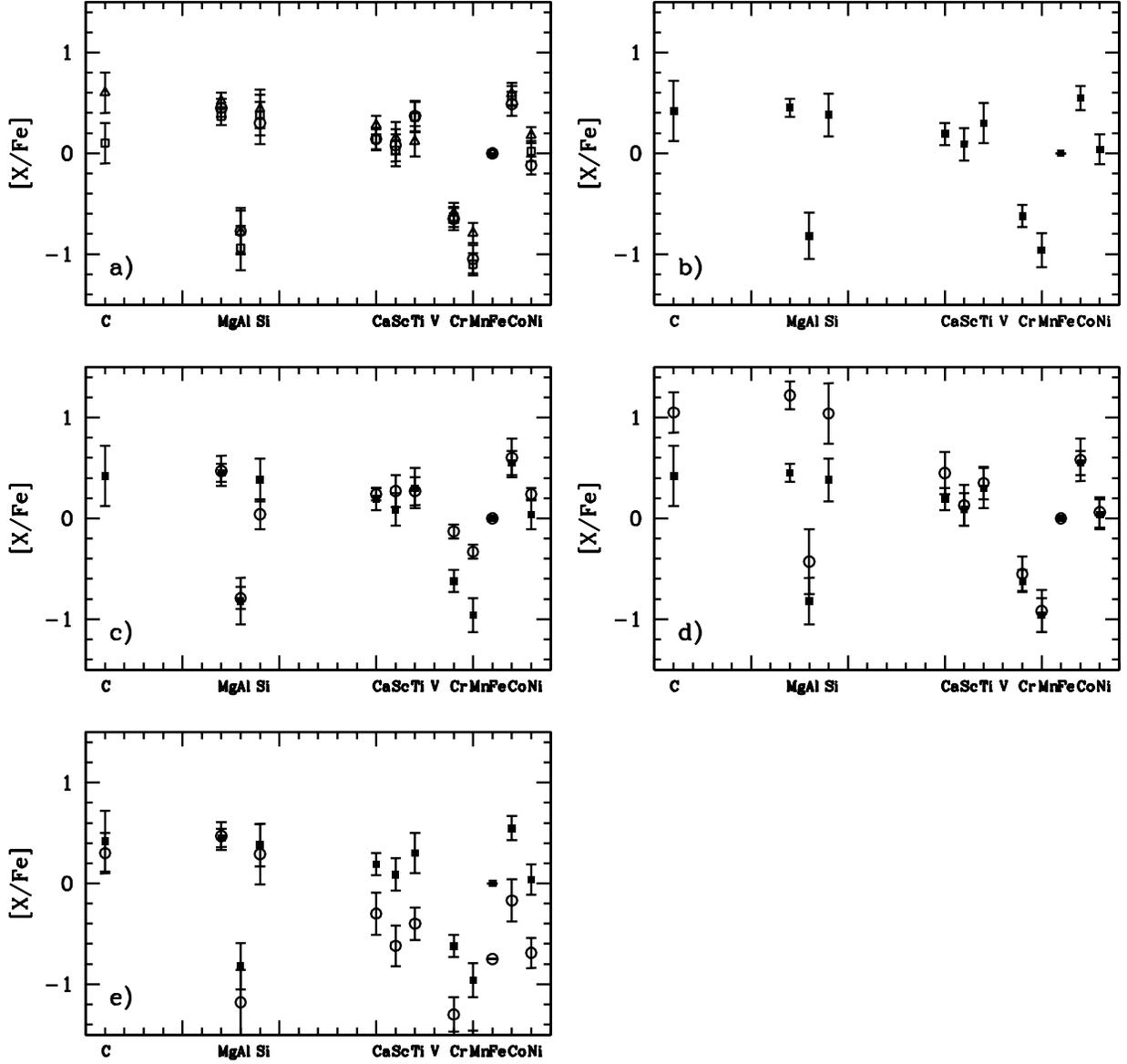} 
\caption{{\em Panel a):} elemental abundances of the three ultra metal poor stars $\rm CD-38^{\rm o}245$ ([Fe/H]=-
3.98) ({\em open circles}), CS 22172-002 ([Fe/H]=-3.61) ({\em open squares}) and CS 22885-096 ([Fe/H]=-3.66) ({\em open triangles}). 
{\em Panel b):} chemical pattern of the AVG star, which has been obtained by averaging the abundances of the elements shown in panel 
a). {\em Panel c):} comparison between $\rm CD-24^{\rm o}17504$ ([Fe/H]=-3.37) ({\em open circles}) and the AVG star ({\em filled 
squares}). {\em Panel d):} comparison between CS 22949-037 ([Fe/H]=- 3.79) ({\em open circles}) and the AVG star ({\em filled 
squares}). {\em Panel e):} same as panel d) but with the abundances of CS 22949-037 ([Fe/H]=- 3.79) ({\em open circles}) shifted 
downward by 0.7 dex.\label{fig01}}
\end{figure}

\section{The observable} The most recent and homogeneous database of surface abundances of extremely metal poor stars available up 
to now is the one published by \cite{NRB01} and consists of five stars of metallicity lower than [Fe/H]=-3.3. Three out of these 
five stars, i.e. $\rm CD-38^{\rm o}245$ ([Fe/H]=-3.98), CS 22172-002 ([Fe/H]=-3.61) and CS 22885-096 ([Fe/H]=-3.66) show a 
remarkably similar pattern (panel a in Figure \ref{fig01}) with the exception of [C/Fe] which shows significative differences 
between CS 22172- 002 and CS 22885-096 (no C abundance is available for $\rm CD-38^{\rm o}245$). Given the close similarity among 
these three stars it is meaningful to define an "average" (hereinafter AVG) star which represents all three of them. The chemical 
pattern of this AVG star is shown in panel b) of the same figure. The comparison between the AVG star and $\rm CD-24^{\rm o}17504$ 
([Fe/H]=-3.37) is shown in panel c). Also this star shows a chemical pattern which closely matches that of the other three stars 
with the exception of the elements Cr and Mn which are significantly more abundant (by a factor three to four) in $\rm CD-24^{\rm 
o}17504$ than in the other three stars. No C abundance determination is available for this star. Panel d) in the same figure shows 
the comparison between the AVG star and CS 22949-037 ([Fe/H]=- 3.79). While there is a clear agreement between this and the AVG star 
for the elements from Ca to Ni, the lighter ones appear to be, in this star, much more abundant (by a factor of five on the mean) 
than in the AVG star. This is clearly shown in panel e) where all the abundances of CS 22949-037 have been shifted downward by 0.7 
dex: all the elements between C and Si now follow a pattern similar to the one shown by the AVG star. This star shows also an 
extremely high [N/Fe] ($\simeq$ 2.7 dex), value which is much larger than the factor of five required to fit the bulk of the 
elements up to Si.  Leaving apart this last star which certainly shows significative differences respect to the other ones, we feel 
confident to say that the other four stars are similar enough to be well represented by the AVG star: therefore in the following we 
will consider this "template" star as the "observable" worth to be compared with the theoretical expectations. 

\section{Theoretical yields} The theoretical yields we will adopt in the following analysis are based on a new set of zero 
metallicity massive stars ranging in mass between $\rm 15~and~80~M_\odot$. These evolutions have been computed with the latest 
version of the FRANEC code \citep{CLS98}, rel. 4.84. This version of the code does not differ significantly from the one adopted by 
\cite{LSC00}. Let us just remind that these stellar models have been evolved with a network including 179 isotopes fully coupled to 
the physical evolution of the stars from the pre main sequence until the central temperature reaches $6~10^9$ K. A quantity which 
plays a major role in determining the final yields of most of the intermediate mass elements is the Carbon abundance left by the 
central He burning as largely discussed by, e.g., \cite{ww93} and \cite{ietal01}. This abundance is determined by the combined 
effects of the $\rm ^{12}C(\alpha,\gamma)^{16}O$ reaction rate and the behavior of the convective core during the latest part of the 
central He burning phase \citep{ietal01}; however, in a classical scenario in which the border of the convective core is determined 
strictly by the Schwarzschild criterion, the final C abundance is mainly controlled by the adopted $\rm ^{12}C(\alpha,\gamma)^{16}O$ 
rate. In order to bracket the possible values for this rate we chose to compute two sets of models, one by adopting the rate 
published by \cite{ca85}, hereinafter set H, and a second one with the rate given by \cite{cf88}, hereinafter set L. For sake of 
completeness Table 1 lists the C mass fraction $\rm X_C$ left by the He burning for all the masses in the two sets of 
models.

\begin{figure}
\epsscale{0.8} 
\plotone{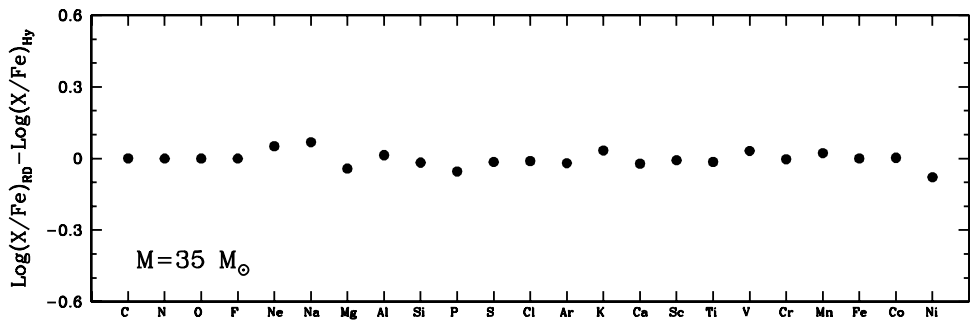} 
\caption{Comparison between the final explosive (X/Fe) computed with two different techniques: radiation dominated 
shock approximation versus hydro code.\label{fig02}}
\end{figure} 

The explosive nucleosynthesis has been computed within the frame of the radiation dominated shock approximation
\citep{ww80,ar96}. More specifically 
it is assumed that, as the shock wave (escaping the iron core with energy E) moves outward in mass through the massive star 
envelope, the pressure inside the shock front is nearly constant and dominated by the radiation pressure. Under this assumption
each zone of the presupernova model is heated up to a temperature $T_{\rm shock}$ given by:
$$
E_{\rm expl}={4\over3} \pi R^{3} a T_{\rm shock}^{4}
$$
where $a$ is the Stefan-Boltzmann constant, $E_{\rm expl}$ is the explosion energy, $T_{\rm shock}$ and $R$ are the temperature behind the shock and its location. 
The shock density can be easily derived by imposing the shock to be mildly strong, i.e.,
$$
\rho_{\rm shock}=f\cdot \rho_{\rm pre-shock}
$$
where $f=4$. By the way let us underline that the final yields would not significantly change even assuming a
strong shock ($f=7$).
The temporal variation of the temperature and density in each mass layer is obtained 
by assuming the matter to expand adiabatically ($T\propto \rho^{\gamma-1}$) at the escape 
velocity ($v_{\rm}=\sqrt{2GM/R}$) and the density to decline following an exponential low, i.e.:
$$
\rho(t)=\rho_{\rm shock}\cdot e^{-t/\tau}
$$
where
$$
\tau=\biggl({{1}\over{\rho}}{{\partial \rho}\over{\partial t}}\biggr)^{-1}={{R}\over{3v}}={{1}\over{\sqrt{24\pi G\rho}}}={446 \over \sqrt{\rho}}
$$
In these computations (unless 
explicitly stated) we assume always a final explosion energy equal to $\rm 1.2\times10^{51}$ erg (1.2 foe) and 
an initial collapse of all the mass zones outside the iron core over a time of 0.5 s \citep{tnh96}.
The chemical evolution has been followed by solving 
the same nuclear network adopted into the presupernova evolutions together to the temporal variation
of temperature and density.

The elemental yields coming from set H have been already 
published by \cite{lc02pasa}, while the ones obtained for the set L are reported in Tables 2-7: for each mass we provide, as we 
usually do, the elemental yields in solar masses for different possible choices of the $\rm ^{56}Ni$ ejected (shown in the first 
row), after all unstable isotopes have been decayed into their stable form.

In addition to the full set of explosions computed within the frame of the radiation dominated shock 
approximation, we have also computed one explosion with an hydro code. In particular we computed the explosion 
of the $\rm 35 M_\odot$ (set H). The hydro code solves the full hydrodynamical equations (included the 
gravitational field) in spherical symmetry and in lagrangean form together to the same nuclear network adopted 
into the presupernova evolutions. In this case the explosion has been induced by imposing the innermost edge 
of the envelope to move like a piston having an initial velocity $V_0$ and following a simple ballistic 
trajectory under the gravitational field of the compact remnant. The present version of this code does not 
take into account the time delay between collapse and explosion yet. The initial velocity $V_0$ has been tuned 
to obtain the ejection of 0.10 $\rm M_\odot$ of $\rm ^{56}Ni$: the corresponding final kinetic energy is 2.86 
foes and the mass cut (i.e. the mass limit which separates the fraction of the star actually ejected in the 
interstellar medium from the fraction which remains locked in the remnant) is located at 2.37 $\rm M_\odot$. 
Also the explosive yields computed with the radiation dominated shock approximation were obtained by imposing 
both the same kinetic energy at infinity and the ejection of $\rm 0.1~M_\odot$ of $\rm ^{56}Ni$. It is 
interesting to note that the mass coordinate which corresponds to this amount of $\rm ^{56}Ni$, i.e. 2.36 $\rm 
M_\odot$, is very close to the one obtained with the hydro code. Figure \ref{fig02} shows that the two sets of 
yields are in excellent agreement, the differences remaining confined well within 0.1 dex. We computed other 
explosions on the same initial model for different final kinetic energies and found always a good agreement 
between the two techniques. The goodness of this comparison obtained for this specific 
model obviously does not guarantee that the same agreement holds for other masses. We plan to address this 
problem in a forthcoming paper.

\begin{figure}
\epsscale{1.0} 
\plotone{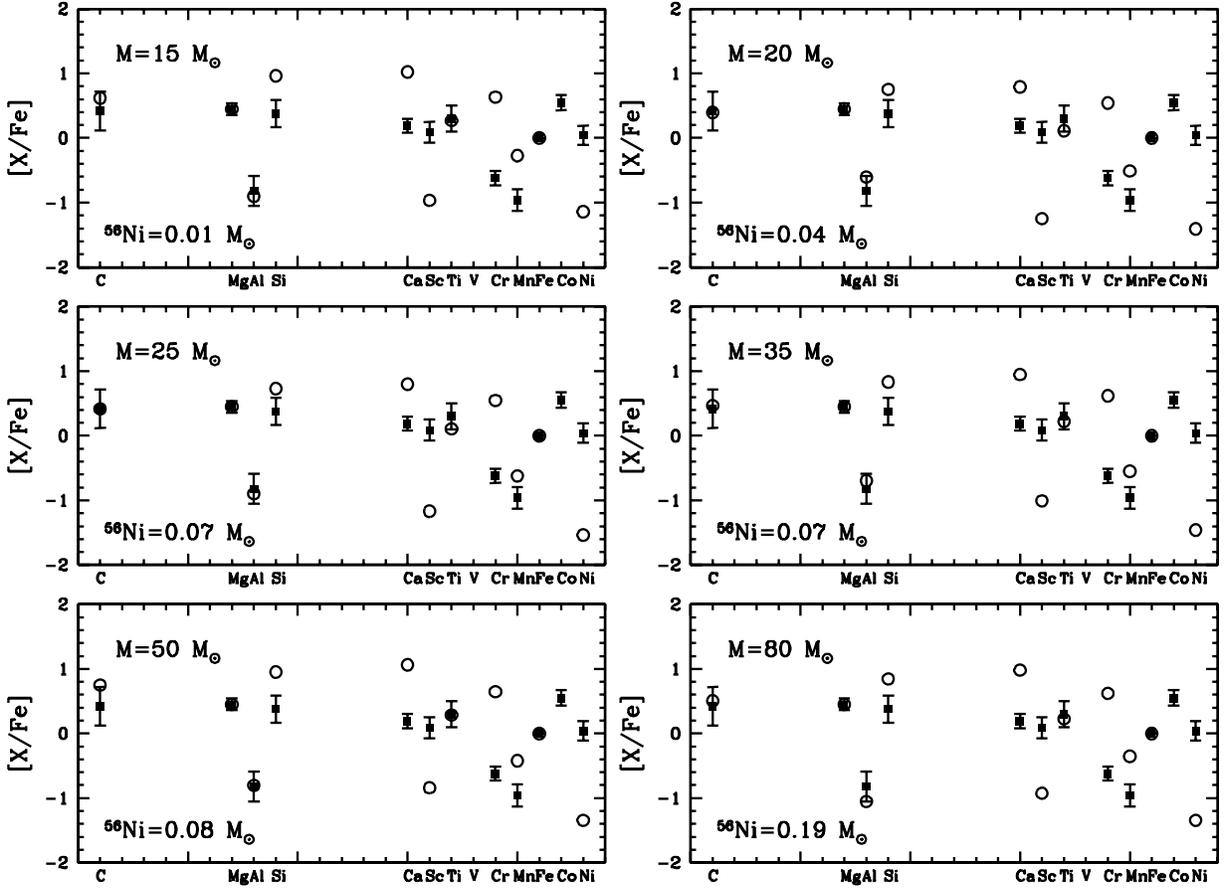} 
\caption{Comparison between the AVG star ({\em filled squares}) and the ejecta of the six masses in our grid ({\em 
open circles}). These models have been computed by adopting the $\rm ^{12}C(\alpha,\gamma)^{16}O$ reaction rate quoted by 
\cite{ca85}. The $\rm ^{56}Ni$ shown in each panel is the one needed to fit the observed [Mg/Fe].\label{fig03}}
\end{figure} 

\section{Fit to the AVG star} 
\begin{figure}
\epsscale{1.0} 
\plotone{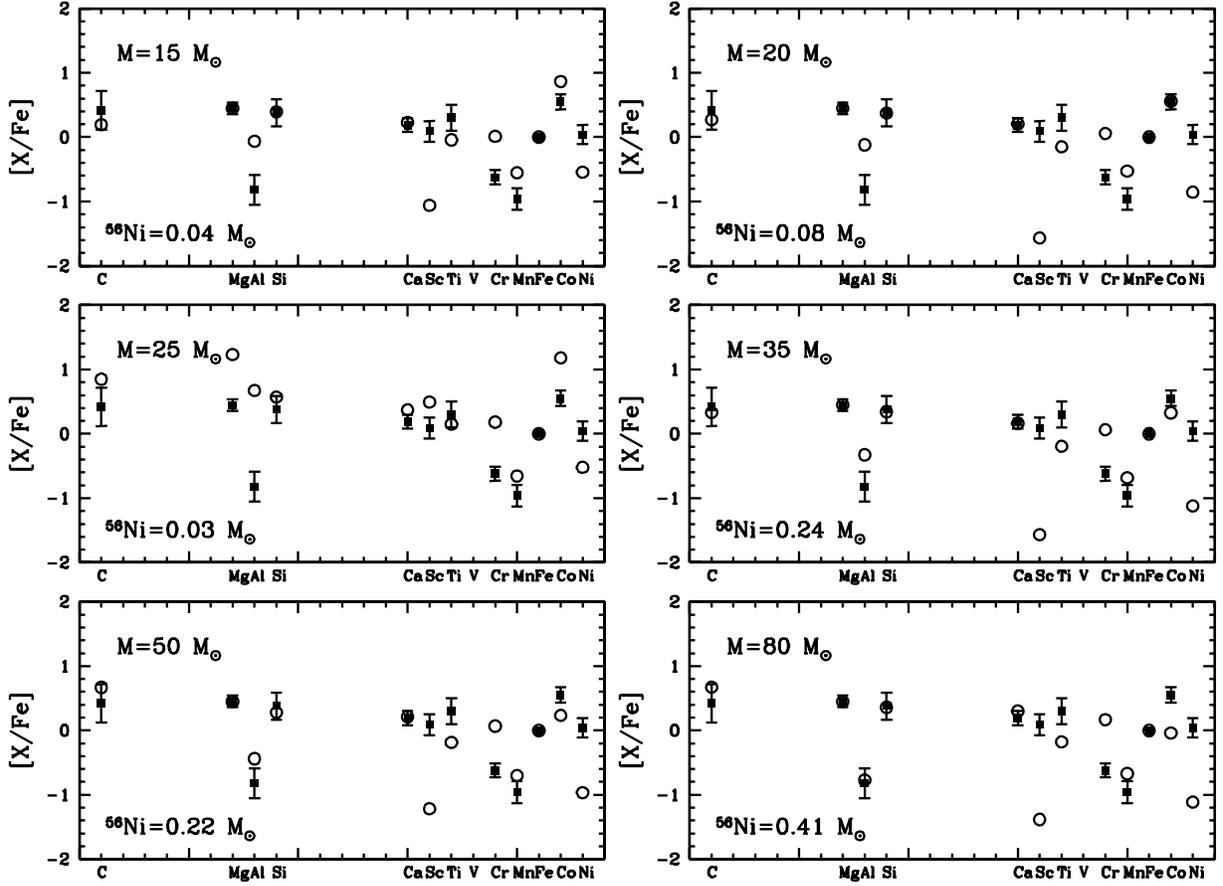} 
\caption{Same as Figure 3 but with models computed by adopting the $\rm ^{12}C(\alpha,\gamma)^{16}O$ reaction rate 
quoted by \cite{cf88}.\label{fig04}}
\end{figure}

Let us discuss the fit to the AVG star under the hypothesis that its elemental ratios are the result 
of a single type II supernova event. The lack of a realistic description of the explosion of a core collapse supernova allows some 
freedom about the choice of the mass cut. Under the hypothesis that each cloud of pristine material is polluted by just a single 
supernova event, a possible way to choose the mass cut is simply that of imposing the fit to a given [X/Fe]. Since Mg is rather well 
measured and it is also produced sufficiently far from the center that it is not significantly affected by the precise location of 
the mass cut, we tentatively chose to fix the mass cut by imposing the fit to the observed [Mg/Fe]. It goes without saying that, 
since the yield of Mg varies with the initial mass, also the mass cut will vary with the mass of the star. The fit to the AVG star 
which is obtained with this choice for the mass cut is shown in Figure \ref{fig03} (for the set H) and in Figure \ref{fig04} (for 
the set L). Figure \ref{fig03} shows that the fit to the two light elements C and Al (Mg is imposed) is remarkably good for all the 
six masses in our grid while it worsens progressively as one moves towards heavier nuclei. Neither Si nor Ca are fitted by any mass 
in our grid so as their internal ratio [Si/Ca]. The fit to the heaviest elements (Sc, Ti, Cr, Mn, Fe, Co and Ni), which are produced 
only by the complete and/or incomplete explosive Si burning, is even worst, with  the exception of Ti which is very well fitted by 
all the models.

\begin{figure}
\epsscale{0.7} 
\plotone{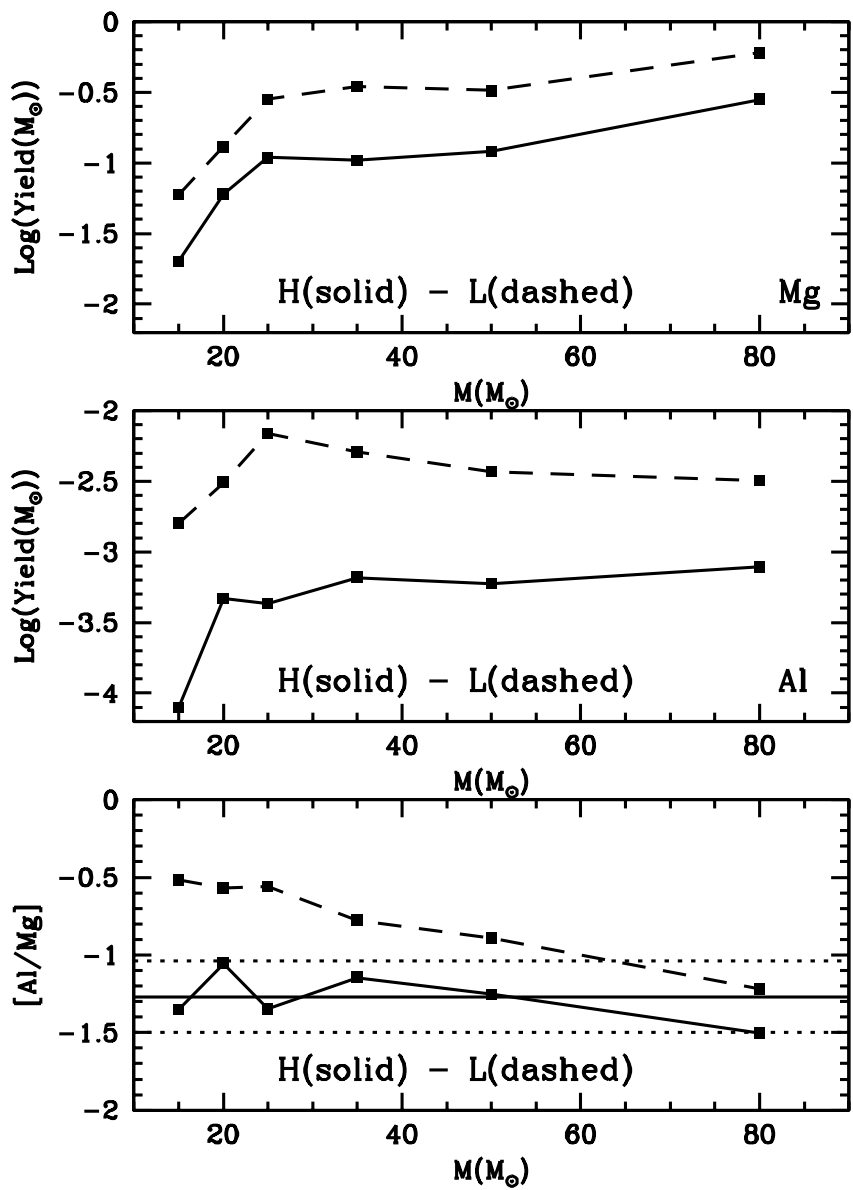} 
\caption{{\em Upper panel}: yield of Mg as a function of the initial mass for both set L (solid line) and set H 
(dashed line). {\em Middle panel}: same as the upper panel but for Al. {\em Lower panel}: [Al/Mg] as a function of the initial mass 
for both set L (solid line) and set H (dashed line). The solid and dotted horizontal lines mark the [Al/Mg] value and the error bar 
of the AVG star.\label{fig05}}
\end{figure}

The fit to the AVG star with the set L (Figure \ref{fig04}) shows a better agreement because in this case both Si and Ca are very 
well reproduced by almost all stellar models. By the way, note that the $25~M_\odot$ does not produce enough $\rm ^{56}Ni$ to fit 
the observed [Mg/Fe]; therefore in this case we simply fixed the mass cut at the border of the Fe core, choice which corresponds to 
the largest possible amount of $\rm ^{56}Ni$. Conversely, [Al/Fe], which was fitted by all the masses of set H, now is fitted only 
by our most massive stellar model, i.e. the $80 M_\odot$. A detailed discussion of the dependence of the yields of the various 
elements on the carbon abundance left by the He burning may be found in \cite{ietal01} and hence here we simply remind that Mg and 
Al scale directly with the Carbon abundance left by the He burning because are both direct products of the Carbon burning. The 
yields of Mg and Al produced by the two sets (H and L) of models are shown in the two upper panels of Figure \ref{fig05}, while the 
trend of [Al/Mg] with the initial mass for the two sets of models is shown in the lowest panel of the same figure. The value of 
[Al/Mg] for the AVG star (-1.27 dex) (solid horizontal line) as well as its error bar (dotted lines) are also shown in Figure 
\ref{fig05}, lowest panel. It is readily evident that while all the models of set H fall within the error bar of the observed 
[Al/Mg], most of the models of set L lie well above the range of compatibility. The adoption of set L changes also the fit to the 
heaviest elements (Sc through Ni) without leading anyway to an acceptable fit. More specifically: [Sc/Fe] remains systematically at 
least 1 dex below the observed value for all the masses in our grid, [Ti/Fe] (which was always well reproduced in the set H) is now 
underestimated by 0.4-0.6 dex, [Cr/Fe] and [Mn/Fe] are both overproduced (by 0.6 and 0.3 dex respectively) by all masses, [Co/Fe] 
shows a trend with the initial mass (in particular it lowers as the initial mass increases) and a good fit is obtained for a mass of 
the order of 20 $\rm M_{\odot}$  while [Ni/Fe] is always underestimated by a factor ranging between 0.6 and 1 dex.

\begin{figure}
\epsscale{1.0} 
\plotone{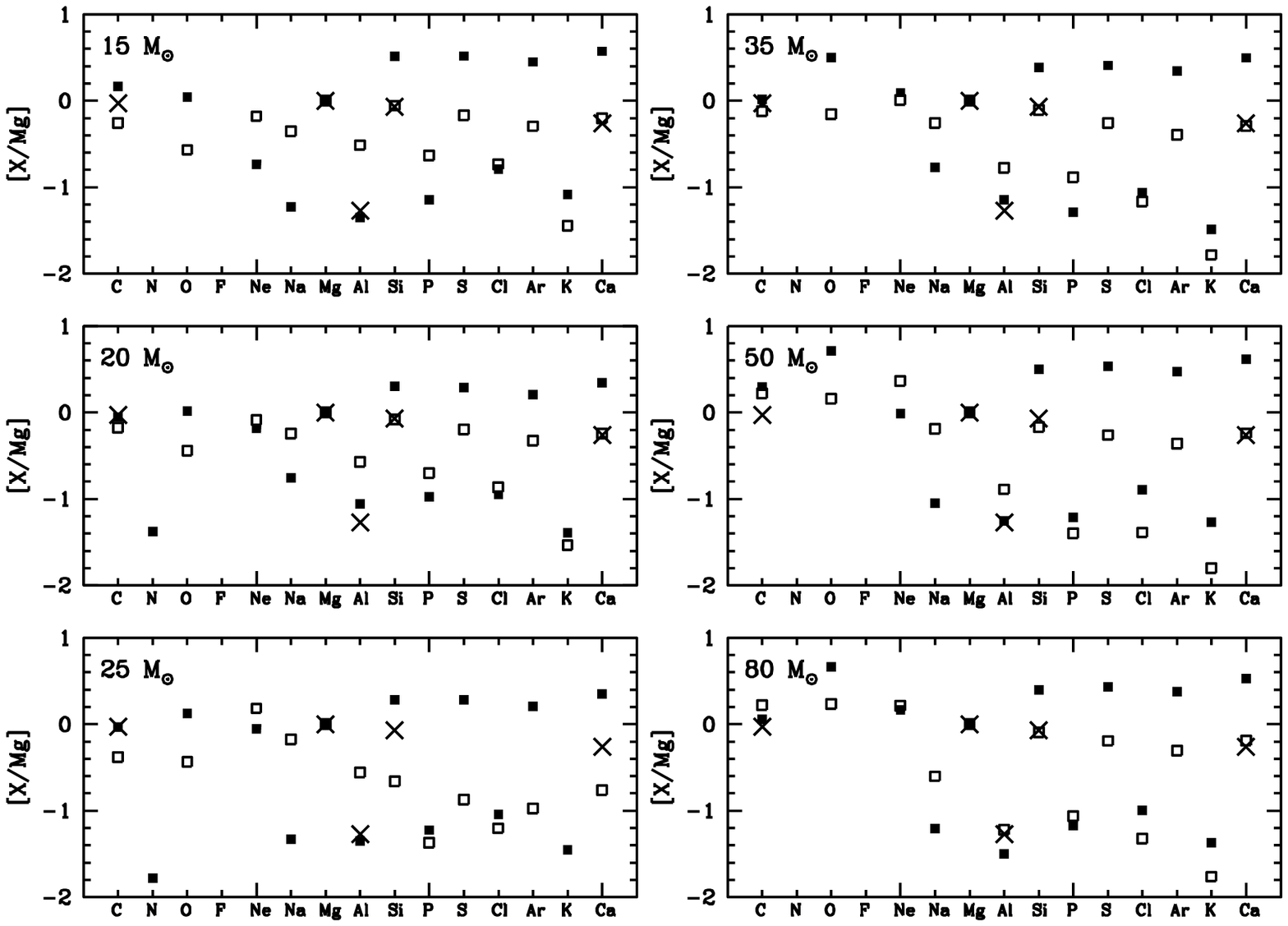} 
\caption{
Comparison between the [X/Mg] obtained for set L (open squares) and set H (filled 
squares). The crosses mark the values obtained for the AVG star. Each panel refers to a specific mass.
\label{fignew}}
\end{figure}

One could argue at this point that there is a contradiction between the present results and those reported by \cite{ietal01} where 
it is stated that a low C abundance (i.e. set H) is necessary to preserve a scaled solar abundances. To clarify this apparent 
contrasting results, we show in Figure \ref{fignew} a comparison between the [X/Mg] obtained with set L (open squares) and set H 
(filled squares), where each panel refers to a specific mass. The comparison refers to the elements which are not significantly 
affected by the mass cut (provided that the fall back is not exceedingly strong). The crosses mark the values obtained for the AVG 
star. \cite{ietal01} computed the evolution of one star (the 25 $\rm M_\odot$) of solar chemical composition and found (among other 
things) that: a) the abundances of the light elements (up to Al and excluding O) scale directly with the C abundance while the 
heaviest ones (up to Ca) scale inversely with C (see fig.16 in \citep{ietal01}) and b) if a 25 $\rm M_\odot$ would be the main 
representative of a generation of massive stars, as suggested also by \cite{ww93}, then a rather low C abundance should be preferred 
to keep the even intermediate mass elements (Mg to Ca) in roughly scaled solar proportions. All the panels in Figure \ref{fignew} 
confirm point a), i.e. that [Si,S,Ar,Ca/Mg] scale inversely with the Carbon abundance left by the central He burning. Also point b) 
is well confirmed by the present computations since also in the case the 25 $\rm M_\odot$ of set H gives a better scaled solar 
abundances for [Si,S,Ar,Ca/Mg] (see Figure \ref{fignew}) than set L. However, the new thing which comes out from these Z=0 models is 
that, for all the other masses in our grid, set L gives a much closer scaled solar distribution than set H does. If the 25 $\rm 
M_\odot$ were not really representative of the ejecta provided by a generation of solar metallicity core collapse supernovae, also in 
that case probably a quite large C abundance at the end of the central He burning should be preferred. In the following, since set L 
provides a much better fit to both [Si/Mg] and [Ca/Mg] than set H does for most of the masses in our grid, we will further analyze 
only models computed with the rate quoted by \cite{cf88} for the $\rm ^{12}C(\alpha,\gamma)^{16}O$.

\begin{figure}
\epsscale{1.0} 
\plotone{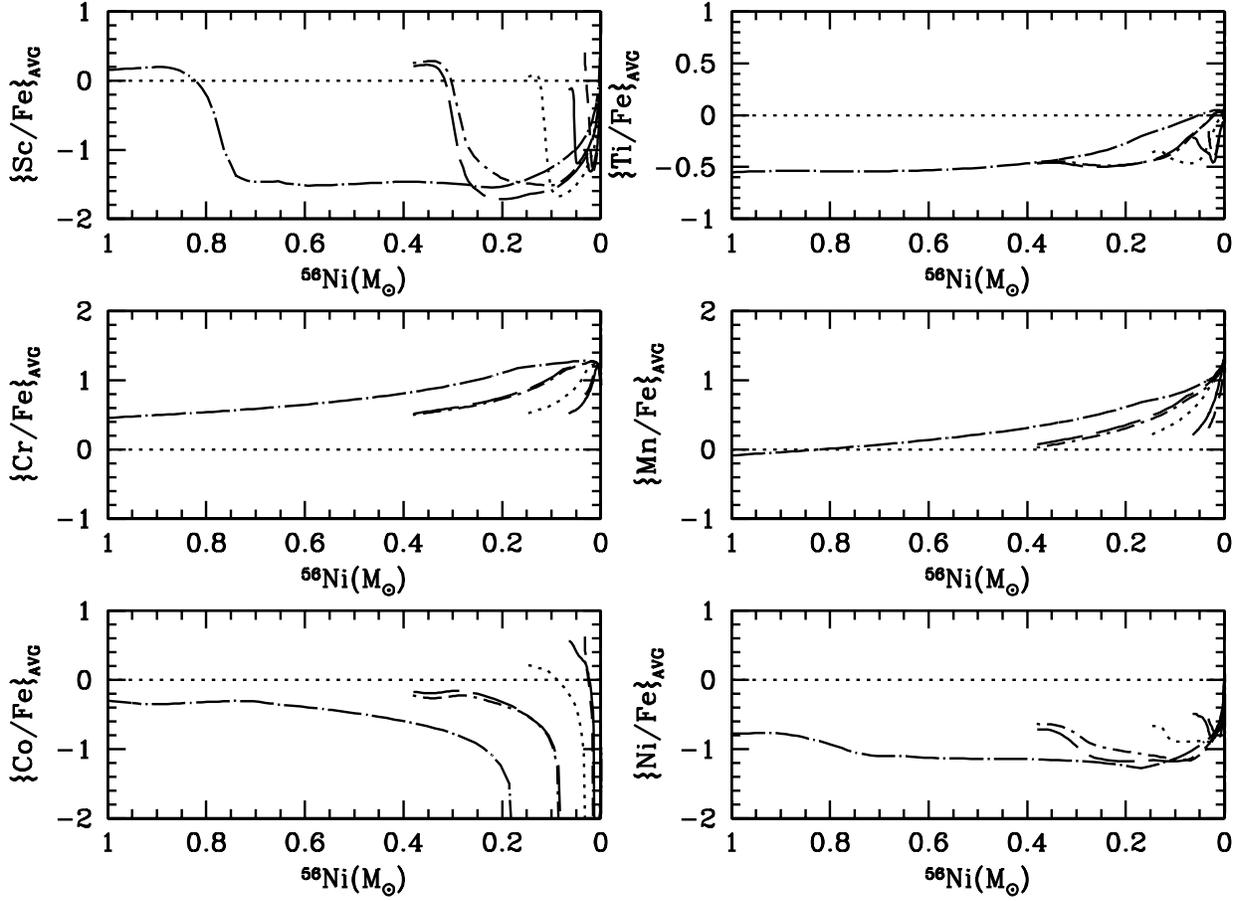} 
\caption{Trends of the [Sc,Ti,Cr,Mn,Co,Ni/Fe], i.e. elements produced by the complete and incomplete explosive Si 
burning, with the amount of Fe ejected. The six lines in each panel refer to: $\rm 15 M_{\odot}$ ({\em solid}), $\rm 20 M_{\odot}$ 
({\em dotted}), $\rm 25 M_{\odot}$ ({\em short dashed}), $\rm 35 M_{\odot}$ ({\em long dashed}), $\rm 50 M_{\odot}$ ({\em dot short 
dashed}), $\rm 80 M_{\odot}$ ({\em dot long dashed}).\label{fig06}}
\end{figure}

\begin{figure}
\epsscale{1.0} 
\plotone{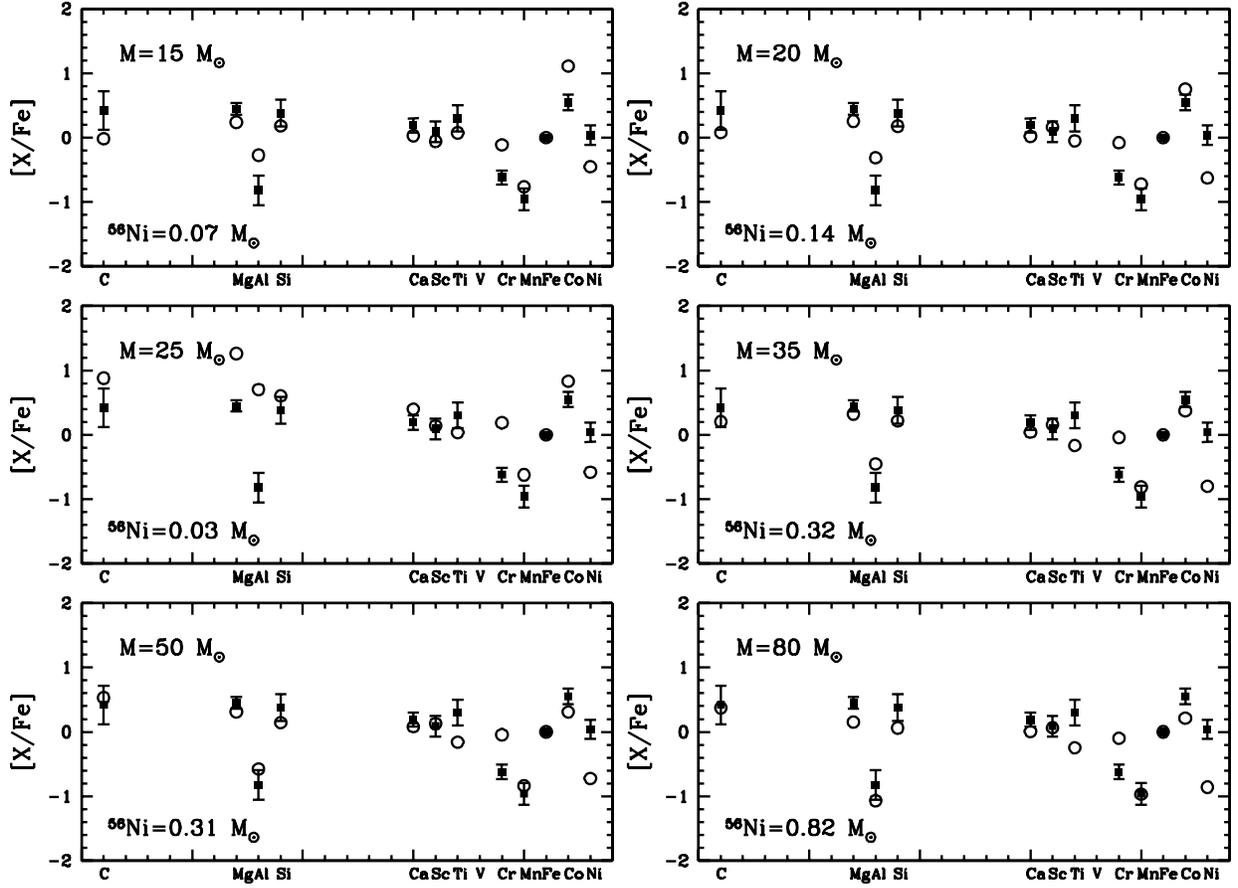} 
\caption{Same as figure 4 but with a choice of the mass cut which minimizes the overall differences between the AVG 
star and the theoretical values.\label{fig07}}
\end{figure}

All the elements between Sc and Ni are produced by either the incomplete and/or the complete explosive Si burning in the deepest 
layers of the star, and hence they are those which are most affected by the adopted mass cut. Hence, the lack of a good fit to these 
elements could be simply due to a wrong choice of the mass cut. In order to show more clearly how the predicted [X/Fe] match the 
corresponding observed values, it is useful to introduce a new abundance ratio: $\{X/Fe\}_{star}$.  This  is simply 
$Log(X_{model}/Fe_{model})-Log(X_{star}/Fe_{star})$, i.e. the difference between the predicted and the observed ratio. This ratio 
differs from the classical "[X/Fe]" simply because in this case the "reference" star is  not the Sun but the star to be fitted. The 
name ${\it star}$ on the lower right corner of this  parameter  clearly identifies  the  "reference star" adopted to compute that 
specific ratio. This is a useful parameter because it readily shows the goodness of a fit (obviously $\{X/Fe\}_{star}=0$ means a 
perfect fit of that star). Figure \ref{fig06} shows how $\rm \{Cr, Mn, Sc, Ti, Co, Ni/Fe\}_{AVG}$ vary with the amount of $\rm 
^{56}Ni$ ejected (i.e. their dependence on the mass cut) for each specific mass. The panels referring to Ti, Ni, and Cr clearly show 
that there is no hope to obtain a good fit to the AVG star for any value of the mass cut, while those referring to the other three 
elements (Sc, Co and Mn), show that (at least for the 20, 35 and 50 $\rm M_\odot$) it exists a value of $\rm ^{56}Ni$ which would 
allow the simultaneous fit to these three elements. Unfortunately this $\rm ^{56}Ni$ abundance is larger than that required to fit 
[Mg/Fe] and hence the fit to the light elements worsens somewhat. The best overall fits to the block of available elements is shown 
in Figure \ref{fig07}. The closest match to the AVG star is obtained with the ejecta of the 35 and 50 $\rm M_\odot$.   

The systematic lack of a good fit to Cr, Ti and Ni clearly indicates that something crucial has been missed in the computation of 
the yields. Though a detailed exploration of all the possible causes for such lack of fit is beyond the scope of the present paper, 
we made some tests by varying both the energy of the shock (between 1 and 3 foes) and the time delay (between 0 and 1 sec) without 
improving appreciably the fits shown above.

\begin{figure}
\epsscale{1.0} 
\plotone{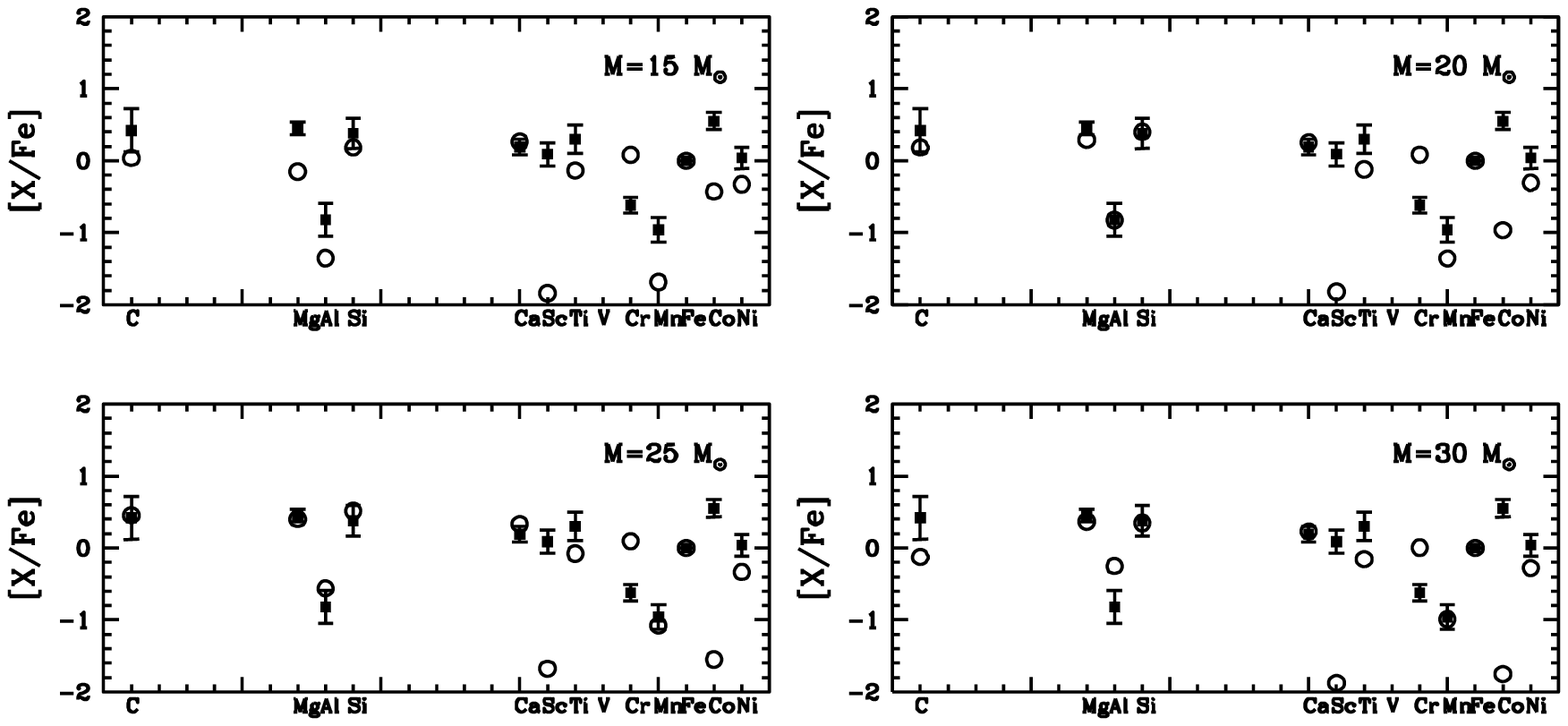} 
\caption{Comparison between the AVG star ({\em filled squares}) and the yields provided by \cite{un02} ({\em open 
circles}).\label{fig08}}
\end{figure}

Alternative yields for zero metal massive stars have been published by \cite{ww95}, hereinafter WW95, and \cite{un02}, hereinafter 
UN02. None of the two papers provides yields as a function of the $\rm ^{56}Ni$ ejected and hence there is no way to impose the fit 
to any of the [X/Fe]. Figure \ref{fig08} shows, for the four masses 15, 20, 25 and 30 $\rm M_{\odot}$, the fits obtained by adopting 
the UN02 yields. The first thing worth noting is that none of the four masses provides an overall good fit to the data. Zooming in 
the various panels, Si and Ca are always rather well fitted (occurrence which should indicate a rather high C abundance) while the 
lighter elements are fitted only in the mass range 20-25 $M_{\odot}$. [Al/Mg] shows a behavior exactly opposite to ours: while we 
find that [Al/Mg] lowers as the initial mass of the star increases, UN02 find that this ratio increases with the mass of the 
star. Also these yields do not allow the fit to Ti, Cr and Ni for any mass in the available range: even more intriguing is the fact 
that the discrepancies are qualitatively very similar to the one we obtain with our yields: in particular, also these yields show 
that [Ti/Fe] and [Ni/Fe] are systematically underproduced with respect to the observed value for all the masses, while [Cr/Fe] is 
always overproduced by the models. The last two elements, Sc and Co, are heavily underproduced by all the UN02 models while Mn is 
well fitted by the two more massive models. 

\begin{figure}
\epsscale{1.0} 
\plotone{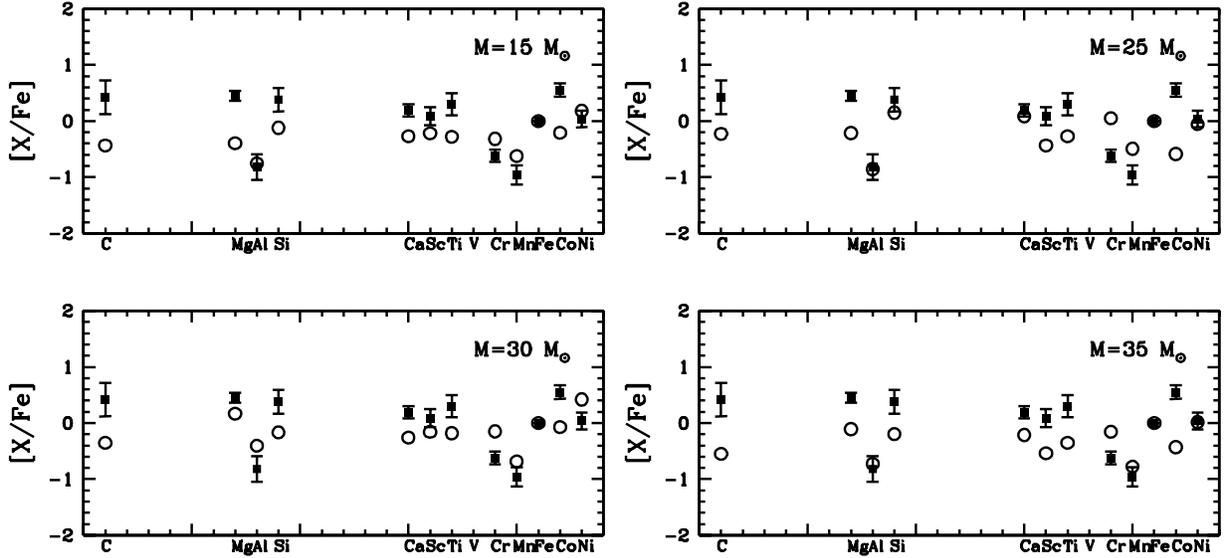} 
\caption{Comparison between the AVG star ({\em filled squares}) and the yields provided by \cite{ww95} ({\em open 
circles}).\label{fig09}}
\end{figure}

Figure \ref{fig09} shows the fit to the AVG star with the yields provided by WW95. The grid is now formed by a 15, 25, 30 and 35 
$\rm M_\odot$. Also in this case none of the models reproduces the data. The element C to Ca are systematically underproduced with 
respect to the observed values and in relative proportions which do not resemble the observations either. [Al/Fe] is vice versa well 
fitted by three out of the four stellar models. Note that also in this case [Al/Mg] is rather small and this could be the hint that 
a rather large C abundance was left by the He burning in these models. The fit to the heaviest elements shows significative 
similarities with those shown above. Once again Sc, Ti and Co are always underproduced with respect to the observed values while Cr 
and Mn are always overproduced. There is however a noticeable difference in these models: [Ni/Fe] is now well fitted or even 
overproduced while neither our models nor those of UN02 are able to produce Ni in sufficient amount.

\begin{figure}
\epsscale{1.0} 
\plotone{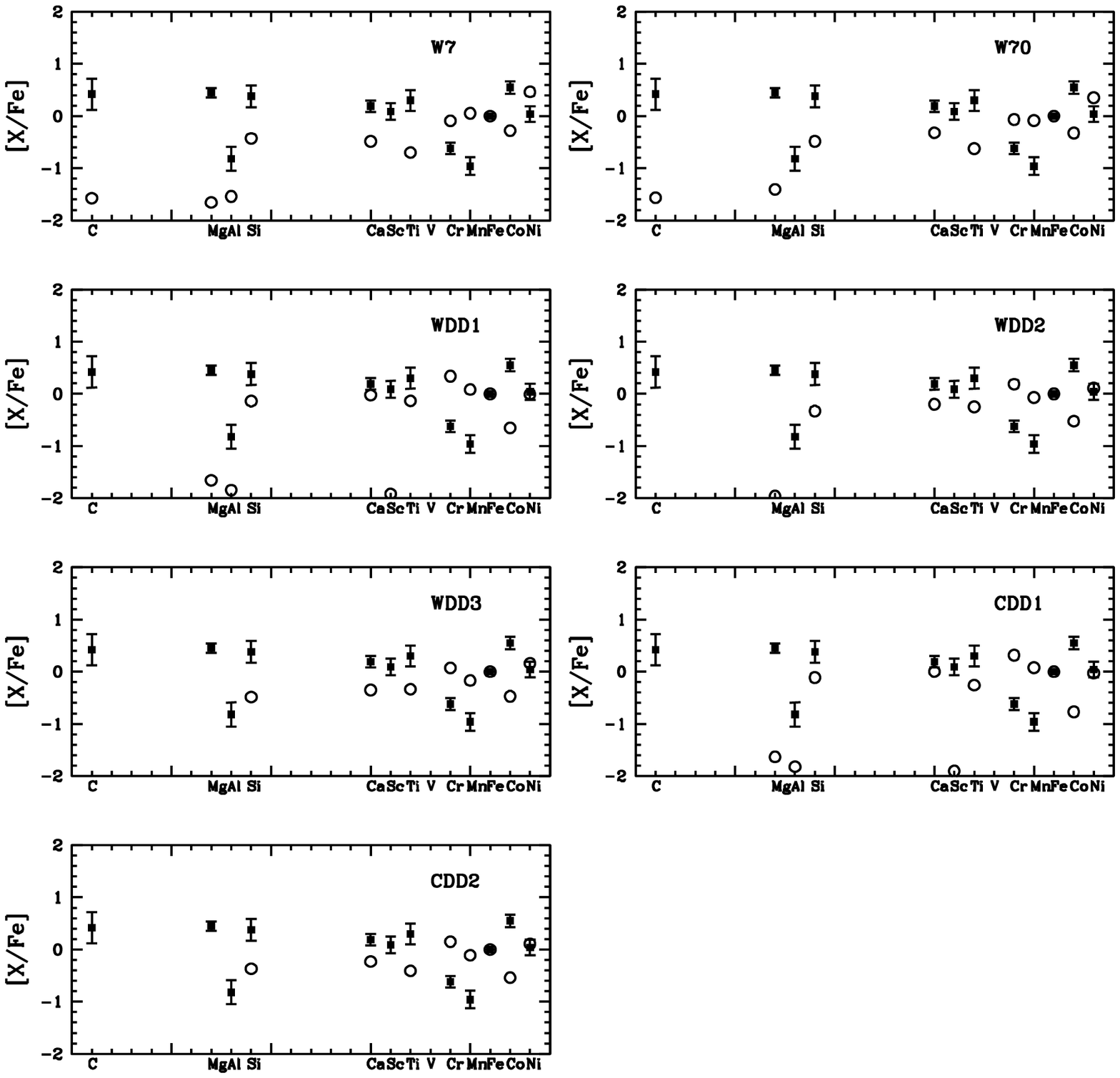} 
\caption{Comparison between the AVG star ({\em filled squares}) and the yields produced by a set of different type Ia 
supernovae provided by \cite{iwa99} ({\em open circles}).\label{fig10}}
\end{figure}

One could argue at this point that the failure of a good fit to the data is simply the consequence of an initial (wrong) assumption: 
i.e. that the observed abundances are due to just the ejecta of a single supernova event and hence that a proper choice of the mass 
function could modify the present analysis. Unfortunately this is not the case for the simple reason that, since no mass (within our 
grid) is able to provide a good fit to the three elements Ti, Cr and Ni, there is obviously no combination of masses which can fit 
them. A last, but not least, possibility is that all these core collapse supernovae fail to reproduce the observed abundances of the 
heaviest elements simply because they do not come from the ejecta of a core collapse supernova but, rather, from the ejecta of a 
type Ia supernova. This would not be impossible, a priori, because the typical lifetime of an intermediate mass star is not much 
different from that of a (low) massive star. To explore such a possibility we tried to fit the AVG star with the yields produced by 
a variety of Type Ia explosions published by \cite{iwa99}. Figure \ref{fig10} shows such a comparison; of course in this case only 
the fit to the heavy elements must be looked at. Also in this case no panel shows a reasonable fit to the data. It is interesting to 
note that also these yields miss the data systematically in the same way as the core collapse supernovae, i.e. Sc, Ti and Co are 
again underproduced while Cr and Mn are systematically overproduced. Ni is the only element which is rather well fitted by most of 
these type Ia explosive models. 

\section{Conclusions}

In the previous sections we have compared the ejecta of primordial core collapse supernovae with the surface chemical composition of 
extremely metal poor stars under the hypothesis that these stars formed in clouds enriched by just the first generation of core 
collapse supernovae. Such a comparison has been done with three different sets of primordial yields: ours, those published by WW95 
and those recently published by UN02. The main general conclusion we feel confident to state is that none of the three sets is able 
to convincingly match the observed abundances. The adoption of a delta function for the mass function cannot be used as an excuse 
for such a defaillance. Though the three sets of yields differ significantly one from the other, all of them show embarrassing 
systematic similarities in missing the fit of the template star. For example, all three sets tend systematically to underproduce Ti 
and Ni and to overproduce Cr. Nonetheless let us stress that our models may account for more or less 2/3 of the available elements.

We have also shown above that the fit to elements like Sc and Co drastically depends on the amount of Carbon left by the He burning
(see Figures \ref{fig03} and \ref{fig04}). 
Such an occurrence could appear surprising because it is not so evident how the Carbon abundance may affect the relative abundances 
of elements produced during the explosion by the complete and incomplete explosive Si burning. Once again we refer the reader to the 
paper by \cite{ietal01} for a comprehensive discussion of such a topic. Here let us simply remind that an increase of the Carbon 
abundance leads to a final mass-radius relation more flattened-out because the contraction of the ONe core is partly slowed down by 
the presence of a very active C-burning shell. As a consequence, the average density in the regions which will experience the 
complete and incomplete explosive Si burning will be lower as well. The net result is that the $\rm \alpha$ rich freeze-out is 
considerably favored and also that the overall amount of $\rm ^{56}Ni$ synthesized is significantly reduced. Both these occurrences 
tend to increase significantly both [Sc/Fe] and [Co/Fe] (for any chosen mass cut location). In addition to that, also the yield of 
Mg increases with the Carbon abundance present in the He exhausted core so that, in order to maintain the ratio $\rm \{Mg/Fe\}_{AVG}$ 
close to zero, the mass cut must be located more internally in order to increase also the Fe yield. It goes without saying that, 
since Sc and Co are mainly produced by the complete explosive Si burning, a deeper mass cut automatically tends to increase their 
outcome. 

The strong dependence of [Co/Fe] on the carbon abundance at the end of central He burning could account for 
the  very low [Co/Fe] obtained by other authors \citep{ww95,un02}. In fact in both computations the interplay 
between the treatment of convection and the adopted $\rm ^{12}C(\alpha,\gamma)^{16}O$ rate leads to a rather 
low carbon abundance at the end of central He burning ($\rm X(^{12}C)\simeq0.2$). Also \cite{Netal99}, who 
tried to explain the trend of increasing Co at low metallicities properly choosing a mass cut-initial stellar 
mass relation, were not able to obtain a sufficiently large [Co/Fe] ratio probably because they used models 
with a small carbon abundance \citep{nh88}. In order to help other groups to compare their results with the 
present ones we report in Table 8 for three selected zones of the  35 $\rm M_\odot$ (set L), where we 
obtain a substantial amount of Sc and Co, the mass coordinate (column 1), the shock temperature (column 2), 
the shock density (column 3), the escape velocity (column 4), the electron mole fraction (column 5), the final 
Co and Ni mass fractions (columns 6 and 7) after the decay of all the unstable isotopes.

Very recently it has been suggested \citep{hw02} that either an observed very low [Al/Mg] ratio as well as observed solar ratios 
among even elements like Si, S, Ar and Ca (e.g. $\rm [Si/Ca]\simeq0$) could be a signature of a primordial generation of very 
massive supernovae ($\rm 260~M_{\odot} > M > 140~M_{\odot}$). Though this is certainly possible, there are some points we think to 
be important to clarify. First of all the production (even in appreciable amounts) of odd elements like Na and Al  {\em does not} 
require any neutron excess because it exists a primary sequence ${\rm ^{12}C(^{12}C,p)^{23}Na(\alpha,\gamma)^{27}Al}$ (where the 
$\alpha$'s also come directly from the C burning: ${\rm ^{12}C(^{12}C,\alpha)^{20}Ne}$) whose efficiency in producing these odd 
elements scales directly with the amount of C present in the He exhausted core. Figure \ref{fig05} clearly shows, e.g., that an $\rm 
80~M_{\odot}$ with a low C abundance produces an [Al/Mg]=- 1.5 dex while a $\rm 15~M_{\odot}$ with a rather large C abundance 
produces an [Al/Mg]=-0.5 dex. Since the amount of C left by the He burning is still subject to severe uncertainties, we think that 
it is not wise to use the ratio of [(light odd elements)/mg] to infer the mass function of the first stellar generation. Also the 
occurrence that the internal ratios among the even elements Si, S, Ar and Ca are solar (e.g. [Si/S], [Si/Ca] and the like) is not a 
typical property of the massive (or very massive) stars but just a property of the incomplete explosive Si burning and of the 
explosive Oxygen burning (also the ejecta of the thermonuclear supernovae provide these elements in roughly solar proportions). Vice 
versa, it is important to underline that also these ratios mildly depend on the amount of C present in the He exhausted 
core \citep{ww93,ietal01}. 

We want eventually stress that since each (actually most) of the elemental ratios observed in the ultra metal poor stars may be 
singularly reproduced, only the simultaneous fit to many (hopefully all) of the observed elemental abundances may lead to a better 
comprehension of the evolution and explosion of the real stars. For this reason we consider still partly unsatisfactory the fits 
shown in Figure \ref{fig07} which miss three out of the eleven available elemental abundances.

Before closing this paper let us anticipate that we have almost completed the upgrade of the FRANEC code with the latest available 
cross sections for both weak processes and strong interactions and that we have extended the nuclear network up to Mo. We therefore 
hope to upgrade shortly also the present results and conclusions.  We want also to finally stress that we barely need also 
homogeneous data concerning other elements like, e.g., O, K and V in order to constrain more vigorously the presupernova evolution 
(Oxygen) and the explosion (Potassium and Vanadium) of the stars which provided the chemical composition of the stars we observe 
today.

\acknowledgements It is a pleasure to thank Sean Ryan, John Norris and Tim Beers for useful discussions. Alessandro Chieffi thanks the 
Astronomical Observatory of Rome and its Director, Prof. Roberto Buonanno, for the generous hospitality in Monteporzio Catone. This 
paper has been supported by the Observatory of Rome. Marco Limongi thanks 
John Lattanzio and Brad Gibson for their very generous hospitality during his visit in
Australia where some of the models presented in this paper have been computed using the VPAC facilities.

\begin{deluxetable}{ccc}
\tablewidth{0pt}
\tablecaption{Carbon abundance left by the central He burning}
\tablehead{
\colhead{Mass($\rm M_\odot)$} & \colhead{Set L} & \colhead{Set H}
}
\startdata
15  & 0.479 & 0.231 \\
20  & 0.439 & 0.196 \\
25  & 0.407 & 0.168 \\
35  & 0.363 & 0.124 \\
50  & 0.319 & 0.092 \\
80  & 0.269 & 0.061 \\
\enddata
\end{deluxetable}

\begin{deluxetable}{lrrrrr}
\tablewidth{0pt}
\tablecaption{Explosive nucleosynthesis of the $\rm 15~M_\odot~~Z=0$ model} 
\tablehead{
\colhead{}        & \colhead{(1)} & \colhead{(2)} & \colhead{(3)} & \colhead{(4)} & \colhead{(5)} 
}
\startdata  
$\rm M(^{56}Ni)$  &      0.0010  &    0.0050  &    0.0100 &     0.0500 &     0.0616   \\
$\rm M_{ejected}$ &        1.71  &      1.69  &      1.68 &       1.61 &       1.58   \\   
$\rm M_{cut}$     &       13.29  &     13.31  &     13.32 &      13.39 &      13.42   \\   
H                 &    7.74E+00  &  7.74E+00  &  7.74E+00 &   7.74E+00 &   7.74E+00   \\
He                &    4.68E+00  &  4.68E+00  &  4.68E+00 &   4.68E+00 &   4.68E+00   \\
C                 &    1.53E-01  &  1.53E-01  &  1.53E-01 &   1.53E-01 &   1.53E-01   \\
N                 &    4.82E-08  &  4.82E-08  &  4.82E-08 &   4.82E-08 &   4.82E-08   \\
O                 &    2.34E-01  &  2.34E-01  &  2.34E-01 &   2.34E-01 &   2.34E-01   \\
F                 &    2.47E-10  &  2.47E-10  &  2.47E-10 &   2.47E-10 &   2.47E-10   \\
Ne                &    1.05E-01  &  1.05E-01  &  1.05E-01 &   1.05E-01 &   1.05E-01   \\
Na                &    1.33E-03  &  1.33E-03  &  1.33E-03 &   1.33E-03 &   1.33E-03   \\
Mg                &    5.95E-02  &  5.95E-02  &  5.95E-02 &   5.95E-02 &   5.95E-02   \\
Al                &    1.60E-03  &  1.60E-03  &  1.60E-03 &   1.60E-03 &   1.60E-03   \\
Si                &    4.69E-02  &  5.51E-02  &  5.63E-02 &   5.63E-02 &   5.63E-02   \\
P                 &    1.71E-04  &  1.71E-04  &  1.71E-04 &   1.71E-04 &   1.71E-04   \\
S                 &    1.88E-02  &  2.42E-02  &  2.54E-02 &   2.54E-02 &   2.54E-02   \\
Cl                &    5.30E-05  &  5.30E-05  &  5.30E-05 &   5.37E-05 &   5.63E-05   \\
Ar                &    2.82E-03  &  3.89E-03  &  4.20E-03 &   4.22E-03 &   4.23E-03   \\
K                 &    1.20E-05  &  1.20E-05  &  1.20E-05 &   1.20E-05 &   1.20E-05   \\
Ca                &    1.85E-03  &  2.85E-03  &  3.27E-03 &   3.40E-03 &   3.44E-03   \\
Sc                &    2.80E-08  &  3.01E-08  &  3.16E-08 &   3.64E-07 &   1.78E-06   \\
Ti                &    5.85E-06  &  2.11E-05  &  3.05E-05 &   1.28E-04 &   1.77E-04   \\
V                 &    6.92E-07  &  1.71E-06  &  2.09E-06 &   2.10E-06 &   2.10E-06   \\
Cr                &    6.97E-05  &  3.08E-04  &  4.94E-04 &   6.44E-04 &   7.12E-04   \\
Mn                &    3.22E-05  &  8.64E-05  &  1.17E-04 &   1.19E-04 &   1.19E-04   \\
Fe                &    1.55E-03  &  5.93E-03  &  1.11E-02 &   5.29E-02 &   6.50E-02   \\
Co                &    8.68E-08  &  9.13E-08  &  9.45E-08 &   1.43E-03 &   2.21E-03   \\
Ni                &    4.99E-05  &  9.39E-05  &  1.30E-04 &   9.99E-04 &   1.32E-03   \\
Cu                &    1.57E-13  &  1.57E-13  &  1.57E-13 &   1.24E-12 &   2.86E-12   \\
Zn                &    5.12E-13  &  5.12E-13  &  5.19E-13 &   9.85E-08 &   1.87E-07   \\
\enddata            
\end{deluxetable}   

\begin{deluxetable}{lrrrrrr}
\tablewidth{0pt}
\tablecaption{Explosive nucleosynthesis of the $\rm 20~M_\odot~~Z=0$ model} 
\tablehead{
\colhead{}        & \colhead{(1)} & \colhead{(2)} & \colhead{(3)} & \colhead{(4)} & \colhead{(5)} & \colhead{(6)} 
}
\startdata  
$\rm M(^{56}Ni)$  &    0.0010 &     0.0050 &     0.0100 &     0.0500 &      0.1000    &    0.1406   \\
$\rm M_{ejected}$ &      2.00 &       1.96 &       1.94 &       1.89 &        1.81    &      1.74   \\
$\rm M_{cut}$     &     18.00 &      18.04 &      18.06 &      18.11 &       18.19    &     18.26   \\
H                 &  9.75E+00 &   9.75E+00 &   9.75E+00 &   9.75E+00 &    9.75E+00    &  9.75E+00   \\
He                &  6.27E+00 &   6.27E+00 &   6.27E+00 &   6.27E+00 &    6.27E+00    &  6.27E+00   \\
C                 &  4.04E-01 &   4.04E-01 &   4.04E-01 &   4.04E-01 &    4.04E-01    &  4.04E-01   \\
N                 &  7.07E-08 &   7.07E-08 &   7.07E-08 &   7.07E-08 &    7.07E-08    &  7.07E-08   \\
O                 &  6.91E-01 &   6.91E-01 &   6.91E-01 &   6.91E-01 &    6.91E-01    &  6.91E-01   \\
F                 &  1.21E-09 &   1.21E-09 &   1.21E-09 &   1.21E-09 &    1.21E-09    &  1.21E-09   \\
Ne                &  2.87E-01 &   2.87E-01 &   2.87E-01 &   2.87E-01 &    2.87E-01    &  2.87E-01   \\
Na                &  3.81E-03 &   3.81E-03 &   3.81E-03 &   3.81E-03 &    3.81E-03    &  3.81E-03   \\
Mg                &  1.31E-01 &   1.31E-01 &   1.31E-01 &   1.31E-01 &    1.31E-01    &  1.31E-01   \\
Al                &  3.10E-03 &   3.10E-03 &   3.10E-03 &   3.10E-03 &    3.10E-03    &  3.10E-03   \\
Si                &  9.26E-02 &   1.09E-01 &   1.15E-01 &   1.18E-01 &    1.18E-01    &  1.18E-01   \\
P                 &  3.21E-04 &   3.21E-04 &   3.21E-04 &   3.21E-04 &    3.21E-04    &  3.22E-04   \\
S                 &  3.52E-02 &   4.54E-02 &   5.01E-02 &   5.28E-02 &    5.28E-02    &  5.28E-02   \\
Cl                &  9.17E-05 &   9.18E-05 &   9.18E-05 &   9.18E-05 &    9.21E-05    &  1.01E-04   \\
Ar                &  5.01E-03 &   6.84E-03 &   7.89E-03 &   8.66E-03 &    8.68E-03    &  8.70E-03   \\
K                 &  2.16E-05 &   2.16E-05 &   2.17E-05 &   2.17E-05 &    2.17E-05    &  2.18E-05   \\
Ca                &  3.18E-03 &   4.74E-03 &   5.84E-03 &   6.96E-03 &    7.06E-03    &  7.17E-03   \\
Sc                &  5.00E-08 &   5.30E-08 &   5.51E-08 &   5.83E-08 &    1.73E-07    &  6.29E-06   \\
Ti                &  7.68E-06 &   2.82E-05 &   4.77E-05 &   8.73E-05 &    1.81E-04    &  2.95E-04   \\
V                 &  9.65E-07 &   2.54E-06 &   3.57E-06 &   4.99E-06 &    4.99E-06    &  4.99E-06   \\
Cr                &  8.08E-05 &   3.66E-04 &   6.69E-04 &   1.35E-03 &    1.48E-03    &  1.64E-03   \\
Mn                &  4.01E-05 &   1.10E-04 &   1.65E-04 &   2.78E-04 &    2.78E-04    &  2.78E-04   \\
Fe                &  1.83E-03 &   6.40E-03 &   1.17E-02 &   5.30E-02 &    1.05E-01    &  1.47E-01   \\
Co                &  1.38E-07 &   1.43E-07 &   1.46E-07 &   2.21E-04 &    1.28E-03    &  2.24E-03   \\
Ni                &  6.56E-05 &   1.14E-04 &   1.48E-04 &   4.34E-04 &    8.67E-04    &  1.99E-03   \\
Cu                &  2.11E-13 &   2.11E-13 &   2.11E-13 &   2.18E-13 &    4.50E-13    &  1.17E-12   \\
Zn                &  7.10E-13 &   7.11E-13 &   7.11E-13 &   4.13E-09 &    5.17E-08    &  1.45E-07   \\
\enddata           
\end{deluxetable}

\begin{deluxetable}{lrrrr}
\tablewidth{0pt}
\tablecaption{Explosive nucleosynthesis of the $\rm 25~M_\odot~~Z=0$ model} 
\tablehead{
\colhead{}        & \colhead{(1)} & \colhead{(2)} & \colhead{(3)} & \colhead{(4)}
}
\startdata  
$\rm M(^{56}Ni)$  &     0.0010  &    0.0050  &    0.0100  &    0.0313     \\
$\rm M_{ejected}$ &       1.66  &      1.63  &      1.63  &      1.57     \\
$\rm M_{cut}$     &      23.34  &     23.37  &     23.37  &     23.43     \\
H                 &   1.15E+01  &  1.15E+01  &  1.15E+01  &  1.15E+01     \\
He                &   7.86E+00  &  7.86E+00  &  7.86E+00  &  7.86E+00     \\
C                 &   5.49E-01  &  5.49E-01  &  5.49E-01  &  5.49E-01     \\
N                 &   6.89E-08  &  6.89E-08  &  6.89E-08  &  6.89E-08     \\
O                 &   1.51E+00  &  1.51E+00  &  1.51E+00  &  1.51E+00     \\
F                 &   3.47E-09  &  3.47E-09  &  3.47E-09  &  3.47E-09     \\
Ne                &   1.15E+00  &  1.15E+00  &  1.15E+00  &  1.15E+00     \\
Na                &   9.63E-03  &  9.63E-03  &  9.63E-03  &  9.63E-03     \\
Mg                &   2.85E-01  &  2.85E-01  &  2.85E-01  &  2.85E-01     \\
Al                &   6.91E-03  &  6.91E-03  &  6.91E-03  &  6.91E-03     \\
Si                &   5.67E-02  &  6.59E-02  &  6.72E-02  &  6.73E-02     \\
P                 &   1.49E-04  &  1.49E-04  &  1.49E-04  &  1.49E-04     \\
S                 &   1.65E-02  &  2.25E-02  &  2.37E-02  &  2.39E-02     \\
Cl                &   3.61E-05  &  3.61E-05  &  3.61E-05  &  4.17E-05     \\
Ar                &   2.55E-03  &  3.68E-03  &  4.04E-03  &  4.12E-03     \\
K                 &   7.28E-06  &  7.29E-06  &  7.30E-06  &  7.34E-06     \\
Ca                &   1.97E-03  &  2.99E-03  &  3.47E-03  &  3.69E-03     \\
Sc                &   2.23E-08  &  2.41E-08  &  2.52E-08  &  3.08E-06     \\
Ti                &   4.66E-06  &  1.94E-05  &  2.98E-05  &  1.04E-04     \\
V                 &   3.14E-07  &  8.96E-07  &  1.10E-06  &  1.14E-06     \\
Cr                &   5.48E-05  &  2.82E-04  &  4.90E-04  &  6.85E-04     \\
Mn                &   1.89E-05  &  5.26E-05  &  6.94E-05  &  7.48E-05     \\
Fe                &   1.23E-03  &  5.42E-03  &  1.05E-02  &  3.23E-02     \\
Co                &   3.50E-08  &  3.77E-08  &  3.89E-08  &  1.29E-03     \\
Ni                &   3.18E-05  &  6.67E-05  &  9.17E-05  &  5.59E-04     \\
Cu                &   2.87E-14  &  2.87E-14  &  2.87E-14  &  1.64E-11     \\
Zn                &   1.29E-13  &  1.29E-13  &  1.29E-13  &  1.82E-07     \\
\enddata           
\end{deluxetable}

\begin{deluxetable}{lrrrrrrrrrr}
\tabletypesize{\scriptsize}
\rotate
\tablewidth{0pt}
\tablecaption{Explosive nucleosynthesis of the $\rm 35~M_\odot~~Z=0$ model} 
\tablehead{
\colhead{}        & \colhead{(1)} & \colhead{(2)} & \colhead{(3)} & \colhead{(4)} & \colhead{(5)} & \colhead{(6)} & \colhead{(7)} & \colhead{(8)} & \colhead{(9)} & \colhead{(10)}
}
\startdata  
$\rm M(^{56}Ni)$  &    0.0010  &    0.0050  &    0.0100  &    0.0500  &    0.1000  &       0.1500   &      0.2000   &     0.2500 &    0.3000  &     0.3685  \\
$\rm M_{ejected}$ &      2.55  &      2.48  &      2.43  &      2.33  &      2.28  &         2.21   &        2.15   &       2.08 &      2.01  &       1.91  \\
$\rm M_{cut}$     &     32.45  &     32.52  &     32.57  &     32.67  &     32.72  &        32.79   &       32.85   &      32.92 &     32.99  &      33.09  \\
H                 &  1.47E+01  &  1.47E+01  &  1.47E+01  &  1.47E+01  &  1.47E+01  &     1.47E+01   &    1.47E+01   &   1.47E+01 &  1.47E+01  &   1.47E+01  \\
He                &  1.07E+01  &  1.07E+01  &  1.07E+01  &  1.07E+01  &  1.07E+01  &     1.07E+01   &    1.07E+01   &   1.07E+01 &  1.07E+01  &   1.07E+01  \\
C                 &  1.23E+00  &  1.23E+00  &  1.23E+00  &  1.23E+00  &  1.23E+00  &     1.23E+00   &    1.23E+00   &   1.23E+00 &  1.23E+00  &   1.23E+00  \\
N                 &  2.24E-04  &  2.24E-04  &  2.24E-04  &  2.24E-04  &  2.24E-04  &     2.24E-04   &    2.24E-04   &   2.24E-04 &  2.24E-04  &   2.24E-04  \\
O                 &  3.57E+00  &  3.57E+00  &  3.57E+00  &  3.57E+00  &  3.57E+00  &     3.57E+00   &    3.57E+00   &   3.57E+00 &  3.57E+00  &   3.57E+00  \\
F                 &  6.39E-07  &  6.39E-07  &  6.39E-07  &  6.39E-07  &  6.39E-07  &     6.39E-07   &    6.39E-07   &   6.39E-07 &  6.39E-07  &   6.39E-07  \\
Ne                &  9.45E-01  &  9.45E-01  &  9.45E-01  &  9.45E-01  &  9.45E-01  &     9.45E-01   &    9.45E-01   &   9.45E-01 &  9.45E-01  &   9.45E-01  \\
Na                &  9.75E-03  &  9.75E-03  &  9.75E-03  &  9.75E-03  &  9.75E-03  &     9.75E-03   &    9.75E-03   &   9.75E-03 &  9.75E-03  &   9.75E-03  \\
Mg                &  3.49E-01  &  3.49E-01  &  3.49E-01  &  3.49E-01  &  3.49E-01  &     3.49E-01   &    3.49E-01   &   3.49E-01 &  3.49E-01  &   3.49E-01  \\
Al                &  5.13E-03  &  5.13E-03  &  5.13E-03  &  5.13E-03  &  5.13E-03  &     5.13E-03   &    5.13E-03   &   5.13E-03 &  5.13E-03  &   5.13E-03  \\
Si                &  2.11E-01  &  2.45E-01  &  2.69E-01  &  2.94E-01  &  2.95E-01  &     2.95E-01   &    2.95E-01   &   2.95E-01 &  2.95E-01  &   2.95E-01  \\
P                 &  5.59E-04  &  5.59E-04  &  5.59E-04  &  5.60E-04  &  5.60E-04  &     5.60E-04   &    5.60E-04   &   5.60E-04 &  5.60E-04  &   5.60E-04  \\
S                 &  6.67E-02  &  8.60E-02  &  1.01E-01  &  1.21E-01  &  1.22E-01  &     1.22E-01   &    1.22E-01   &   1.22E-01 &  1.22E-01  &   1.22E-01  \\
Cl                &  1.22E-04  &  1.23E-04  &  1.23E-04  &  1.23E-04  &  1.23E-04  &     1.23E-04   &    1.23E-04   &   1.23E-04 &  1.32E-04  &   1.46E-04  \\
Ar                &  8.52E-03  &  1.18E-02  &  1.44E-02  &  1.93E-02  &  1.97E-02  &     1.97E-02   &    1.98E-02   &   1.98E-02 &  1.98E-02  &   1.98E-02  \\
K                 &  3.23E-05  &  3.24E-05  &  3.25E-05  &  3.25E-05  &  3.25E-05  &     3.25E-05   &    3.25E-05   &   3.25E-05 &  3.26E-05  &   3.28E-05  \\
Ca                &  5.61E-03  &  8.25E-03  &  1.05E-02  &  1.60E-02  &  1.68E-02  &     1.68E-02   &    1.69E-02   &   1.69E-02 &  1.70E-02  &   1.71E-02  \\
Sc                &  7.65E-08  &  8.11E-08  &  8.45E-08  &  9.29E-08  &  1.02E-07  &     1.28E-07   &    1.51E-07   &   2.83E-07 &  7.89E-06  &   2.29E-05  \\
Ti                &  8.65E-06  &  3.24E-05  &  6.00E-05  &  1.77E-04  &  2.16E-04  &     2.56E-04   &    3.10E-04   &   3.77E-04 &  4.73E-04  &   5.93E-04  \\
V                 &  1.07E-06  &  2.91E-06  &  4.60E-06  &  8.54E-06  &  9.45E-06  &     9.45E-06   &    9.45E-06   &   9.45E-06 &  9.45E-06  &   9.45E-06  \\
Cr                &  8.57E-05  &  3.91E-04  &  7.59E-04  &  2.84E-03  &  3.67E-03  &     3.72E-03   &    3.80E-03   &   3.89E-03 &  4.03E-03  &   4.19E-03  \\
Mn                &  4.38E-05  &  1.20E-04  &  1.93E-04  &  4.31E-04  &  5.16E-04  &     5.16E-04   &    5.16E-04   &   5.16E-04 &  5.16E-04  &   5.16E-04  \\
Fe                &  2.01E-03  &  6.70E-03  &  1.22E-02  &  5.33E-02  &  1.04E-01  &     1.56E-01   &    2.07E-01   &   2.59E-01 &  3.10E-01  &   3.80E-01  \\
Co                &  1.46E-07  &  1.51E-07  &  1.55E-07  &  1.61E-07  &  1.47E-04  &     4.82E-04   &    9.84E-04   &   1.58E-03 &  1.97E-03  &   2.40E-03  \\
Ni                &  7.57E-05  &  1.27E-04  &  1.70E-04  &  2.87E-04  &  4.41E-04  &     6.67E-04   &    8.73E-04   &   1.21E-03 &  2.49E-03  &   4.61E-03  \\
Cu                &  3.14E-11  &  3.14E-11  &  3.14E-11  &  3.14E-11  &  3.14E-11  &     3.14E-11   &    3.14E-11   &   3.15E-11 &  3.15E-11  &   3.17E-11  \\
Zn                &  9.83E-10  &  9.83E-10  &  9.83E-10  &  9.83E-10  &  2.93E-09  &     1.11E-08   &    2.54E-08   &   4.65E-08 &  7.33E-08  &   1.19E-07  \\
\enddata           
\end{deluxetable}

\begin{deluxetable}{lrrrrrrrrrr}
\tabletypesize{\scriptsize}
\rotate
\tablewidth{0pt}
\tablecaption{Explosive nucleosynthesis of the $\rm 50~M_\odot~~Z=0$ model} 
\tablehead{
\colhead{}        & \colhead{(1)} & \colhead{(2)} & \colhead{(3)} & \colhead{(4)} & \colhead{(5)} & \colhead{(6)} & \colhead{(7)} & \colhead{(8)} & \colhead{(9)} & \colhead{(10)}
}
\startdata  
$\rm M(^{56}Ni)$  &     0.0010  &    0.0050 &     0.0100 &     0.0500  &      0.1000  &      0.1500  &      0.2000 &       0.2500 &       0.3000   &      0.3709   \\
$\rm M_{ejected}$ &       2.61  &      2.55 &       2.50 &       2.39  &        2.34  &        2.28  &        2.21 &         2.14 &         2.06   &        1.96   \\
$\rm M_{cut}$     &      47.39  &     47.45 &      47.50 &      47.61  &       47.66  &       47.72  &       47.79 &        47.86 &        47.94   &       48.04   \\
H                 &   1.94E+01  &  1.94E+01 &   1.94E+01 &   1.94E+01  &    1.94E+01  &    1.94E+01  &    1.94E+01 &     1.94E+01 &     1.94E+01   &    1.94E+01   \\
He                &   1.53E+01  &  1.53E+01 &   1.53E+01 &   1.53E+01  &    1.53E+01  &    1.53E+01  &    1.53E+01 &     1.53E+01 &     1.53E+01   &    1.53E+01   \\
C                 &   2.51E+00  &  2.51E+00 &   2.51E+00 &   2.51E+00  &    2.51E+00  &    2.51E+00  &    2.51E+00 &     2.51E+00 &     2.51E+00   &    2.51E+00   \\
N                 &   1.12E-06  &  1.12E-06 &   1.12E-06 &   1.12E-06  &    1.12E-06  &    1.12E-06  &    1.12E-06 &     1.12E-06 &     1.12E-06   &    1.12E-06   \\
O                 &   6.88E+00  &  6.88E+00 &   6.88E+00 &   6.88E+00  &    6.88E+00  &    6.88E+00  &    6.88E+00 &     6.88E+00 &     6.88E+00   &    6.88E+00   \\
F                 &   4.33E-09  &  4.33E-09 &   4.33E-09 &   4.33E-09  &    4.33E-09  &    4.33E-09  &    4.33E-09 &     4.33E-09 &     4.33E-09   &    4.33E-09   \\
Ne                &   2.01E+00  &  2.01E+00 &   2.01E+00 &   2.01E+00  &    2.01E+00  &    2.01E+00  &    2.01E+00 &     2.01E+00 &     2.01E+00   &    2.01E+00   \\
Na                &   1.07E-02  &  1.07E-02 &   1.07E-02 &   1.07E-02  &    1.07E-02  &    1.07E-02  &    1.07E-02 &     1.07E-02 &     1.07E-02   &    1.07E-02   \\
Mg                &   3.26E-01  &  3.26E-01 &   3.26E-01 &   3.26E-01  &    3.26E-01  &    3.26E-01  &    3.26E-01 &     3.26E-01 &     3.26E-01   &    3.26E-01   \\
Al                &   3.69E-03  &  3.69E-03 &   3.69E-03 &   3.69E-03  &    3.69E-03  &    3.69E-03  &    3.69E-03 &     3.69E-03 &     3.69E-03   &    3.69E-03   \\
Si                &   1.55E-01  &  1.84E-01 &   2.08E-01 &   2.37E-01  &    2.37E-01  &    2.37E-01  &    2.37E-01 &     2.37E-01 &     2.37E-01   &    2.37E-01   \\
P                 &   1.60E-04  &  1.60E-04 &   1.60E-04 &   1.61E-04  &    1.61E-04  &    1.61E-04  &    1.61E-04 &     1.61E-04 &     1.61E-04   &    1.61E-04   \\
S                 &   5.79E-02  &  7.59E-02 &   9.03E-02 &   1.13E-01  &    1.13E-01  &    1.13E-01  &    1.13E-01 &     1.13E-01 &     1.13E-01   &    1.13E-01   \\
Cl                &   6.82E-05  &  6.83E-05 &   6.84E-05 &   6.85E-05  &    6.86E-05  &    6.87E-05  &    6.90E-05 &     7.00E-05 &     8.35E-05   &    9.53E-05   \\
Ar                &   8.77E-03  &  1.20E-02 &   1.46E-02 &   1.97E-02  &    2.00E-02  &    2.00E-02  &    2.00E-02 &     2.00E-02 &     2.00E-02   &    2.00E-02   \\
K                 &   2.92E-05  &  2.93E-05 &   2.93E-05 &   2.94E-05  &    2.94E-05  &    2.94E-05  &    2.94E-05 &     2.94E-05 &     2.96E-05   &    2.97E-05   \\
Ca                &   6.35E-03  &  9.15E-03 &   1.13E-02 &   1.70E-02  &    1.76E-02  &    1.77E-02  &    1.77E-02 &     1.78E-02 &     1.79E-02   &    1.80E-02   \\
Sc                &   8.58E-08  &  9.05E-08 &   9.36E-08 &   1.02E-07  &    1.20E-07  &    1.91E-07  &    3.21E-07 &     7.88E-07 &     1.45E-05   &    2.56E-05   \\
Ti                &   8.45E-06  &  3.07E-05 &   5.77E-05 &   1.73E-04  &    2.05E-04  &    2.52E-04  &    3.12E-04 &     3.86E-04 &     4.93E-04   &    6.00E-04   \\
V                 &   7.87E-07  &  2.15E-06 &   3.57E-06 &   7.36E-06  &    8.00E-06  &    8.00E-06  &    8.00E-06 &     8.00E-06 &     8.00E-06   &    8.00E-06   \\
Cr                &   7.42E-05  &  3.59E-04 &   7.19E-04 &   2.77E-03  &    3.46E-03  &    3.53E-03  &    3.61E-03 &     3.71E-03 &     3.86E-03   &    4.01E-03   \\
Mn                &   3.54E-05  &  9.71E-05 &   1.61E-04 &   3.89E-04  &    4.64E-04  &    4.64E-04  &    4.64E-04 &     4.64E-04 &     4.64E-04   &    4.64E-04   \\
Fe                &   1.73E-03  &  6.22E-03 &   1.16E-02 &   5.27E-02  &    1.04E-01  &    1.55E-01  &    2.06E-01 &     2.58E-01 &     3.08E-01   &    3.81E-01   \\
Co                &   1.21E-07  &  1.25E-07 &   1.28E-07 &   1.34E-07  &    1.33E-04  &    4.65E-04  &    8.63E-04 &     1.37E-03 &     1.65E-03   &    2.12E-03   \\
Ni                &   6.31E-05  &  1.05E-04 &   1.42E-04 &   2.64E-04  &    4.67E-04  &    8.05E-04  &    1.23E-03 &     1.82E-03 &     3.49E-03   &    5.58E-03   \\
Cu                &   9.41E-14  &  9.41E-14 &   9.41E-14 &   9.41E-14  &    9.54E-14  &    1.07E-13  &    1.28E-13 &     1.77E-13 &     2.27E-13   &    9.00E-13   \\
Zn                &   4.12E-13  &  4.12E-13 &   4.12E-13 &   4.12E-13  &    1.90E-09  &    1.14E-08  &    2.52E-08 &     4.72E-08 &     7.33E-08   &    1.39E-07   \\
\enddata           
\end{deluxetable}

\begin{deluxetable}{lrrrrrrrrrr}
\tabletypesize{\scriptsize}
\rotate
\tablewidth{0pt}
\tablecaption{Explosive nucleosynthesis of the $\rm 80~M_\odot~~Z=0$ model} 
\tablehead{
\colhead{}        & \colhead{(1)} & \colhead{(2)} & \colhead{(3)} & \colhead{(4)} & \colhead{(5)} & \colhead{(6)} & \colhead{(7)} & \colhead{(8)} & \colhead{(9)} & \colhead{(10)}
}
\startdata  
$\rm M(^{56}Ni)$  &      0.0010  &    0.0050  &    0.0100  &    0.0500  &      0.1000  &       0.1500 &       0.2000  &     0.2500  &      0.3000   &     0.9921  \\
$\rm M_{ejected}$ &        3.80  &      3.72  &      3.65  &      3.42  &        3.32  &         3.25 &         3.20  &       3.14  &        3.08   &       2.21  \\
$\rm M_{cut}$     &       76.20  &     76.28  &     76.35  &     76.58  &       76.68  &        76.75 &        76.80  &      76.86  &       76.92   &      77.79  \\
H                 &    2.68E+01  &  2.68E+01  &  2.68E+01  &  2.68E+01  &    2.68E+01  &     2.68E+01 &     2.68E+01  &   2.68E+01  &    2.68E+01   &   2.68E+01  \\
He                &    2.47E+01  &  2.47E+01  &  2.47E+01  &  2.47E+01  &    2.47E+01  &     2.47E+01 &     2.47E+01  &   2.47E+01  &    2.47E+01   &   2.47E+01  \\
C                 &    4.67E+00  &  4.67E+00  &  4.67E+00  &  4.67E+00  &    4.67E+00  &     4.67E+00 &     4.67E+00  &   4.67E+00  &    4.67E+00   &   4.67E+00  \\
N                 &    1.98E-06  &  1.98E-06  &  1.98E-06  &  1.98E-06  &    1.98E-06  &     1.98E-06 &     1.98E-06  &   1.98E-06  &    1.98E-06   &   1.98E-06  \\
O                 &    1.50E+01  &  1.50E+01  &  1.50E+01  &  1.50E+01  &    1.50E+01  &     1.50E+01 &     1.50E+01  &   1.50E+01  &    1.50E+01   &   1.50E+01  \\
F                 &    4.09E-09  &  4.09E-09  &  4.09E-09  &  4.09E-09  &    4.09E-09  &     4.09E-09 &     4.09E-09  &   4.09E-09  &    4.09E-09   &   4.09E-09  \\
Ne                &    2.63E+00  &  2.63E+00  &  2.63E+00  &  2.63E+00  &    2.63E+00  &     2.63E+00 &     2.63E+00  &   2.63E+00  &    2.63E+00   &   2.63E+00  \\
Na                &    7.62E-03  &  7.62E-03  &  7.62E-03  &  7.62E-03  &    7.62E-03  &     7.62E-03 &     7.62E-03  &   7.62E-03  &    7.62E-03   &   7.62E-03  \\
Mg                &    6.04E-01  &  6.04E-01  &  6.04E-01  &  6.04E-01  &    6.04E-01  &     6.04E-01 &     6.04E-01  &   6.04E-01  &    6.04E-01   &   6.04E-01  \\
Al                &    3.19E-03  &  3.19E-03  &  3.19E-03  &  3.19E-03  &    3.19E-03  &     3.19E-03 &     3.19E-03  &   3.19E-03  &    3.19E-03   &   3.19E-03  \\
Si                &    3.35E-01  &  3.75E-01  &  4.07E-01  &  5.05E-01  &    5.24E-01  &     5.28E-01 &     5.28E-01  &   5.28E-01  &    5.28E-01   &   5.28E-01  \\
P                 &    6.47E-04  &  6.47E-04  &  6.47E-04  &  6.48E-04  &    6.49E-04  &     6.49E-04 &     6.49E-04  &   6.49E-04  &    6.49E-04   &   6.50E-04  \\
S                 &    1.16E-01  &  1.42E-01  &  1.61E-01  &  2.28E-01  &    2.46E-01  &     2.51E-01 &     2.51E-01  &   2.51E-01  &    2.51E-01   &   2.51E-01  \\
Cl                &    1.46E-04  &  1.46E-04  &  1.46E-04  &  1.46E-04  &    1.46E-04  &     1.46E-04 &     1.46E-04  &   1.47E-04  &    1.47E-04   &   1.85E-04  \\
Ar                &    1.60E-02  &  2.09E-02  &  2.41E-02  &  3.74E-02  &    4.21E-02  &     4.37E-02 &     4.39E-02  &   4.39E-02  &    4.39E-02   &   4.40E-02  \\
K                 &    5.86E-05  &  5.87E-05  &  5.87E-05  &  5.89E-05  &    5.89E-05  &     5.89E-05 &     5.89E-05  &   5.89E-05  &    5.89E-05   &   5.98E-05  \\
Ca                &    1.17E-02  &  1.61E-02  &  1.87E-02  &  3.13E-02  &    3.72E-02  &     3.96E-02 &     4.00E-02  &   4.00E-02  &    4.01E-02   &   4.07E-02  \\
Sc                &    1.69E-07  &  1.77E-07  &  1.80E-07  &  1.95E-07  &    2.03E-07  &     2.07E-07 &     2.22E-07  &   2.85E-07  &    3.71E-07   &   5.27E-05  \\
Ti                &    1.11E-05  &  3.34E-05  &  6.12E-05  &  2.40E-04  &    3.77E-04  &     4.62E-04 &     4.84E-04  &   5.09E-04  &    5.45E-04   &   1.28E-03  \\
V                 &    9.90E-07  &  2.38E-06  &  3.82E-06  &  1.04E-05  &    1.37E-05  &     1.56E-05 &     1.60E-05  &   1.60E-05  &    1.60E-05   &   1.60E-05  \\
Cr                &    8.14E-05  &  3.52E-04  &  7.15E-04  &  3.31E-03  &    5.84E-03  &     7.85E-03 &     8.31E-03  &   8.34E-03  &    8.39E-03   &   9.41E-03  \\
Mn                &    4.37E-05  &  1.10E-04  &  1.74E-04  &  5.03E-04  &    7.17E-04  &     8.80E-04 &     9.20E-04  &   9.20E-04  &    9.20E-04   &   9.20E-04  \\
Fe                &    2.13E-03  &  6.70E-03  &  1.22E-02  &  5.39E-02  &    1.05E-01  &     1.55E-01 &     2.06E-01  &   2.57E-01  &    3.09E-01   &   1.02E+00  \\
Co                &    2.05E-07  &  2.11E-07  &  2.14E-07  &  2.24E-07  &    2.28E-07  &     2.30E-07 &     1.18E-04  &   2.97E-04  &    4.99E-04   &   4.74E-03  \\
Ni                &    9.57E-05  &  1.47E-04  &  1.86E-04  &  3.61E-04  &    4.60E-04  &     5.38E-04 &     7.45E-04  &   1.03E-03  &    1.30E-03   &   1.04E-02  \\
Cu                &    1.32E-13  &  1.32E-13  &  1.32E-13  &  1.32E-13  &    1.32E-13  &     1.32E-13 &     1.32E-13  &   1.34E-13  &    1.38E-13   &   4.81E-13  \\
Zn                &    6.26E-13  &  6.26E-13  &  6.26E-13  &  6.27E-13  &    6.27E-13  &     6.27E-13 &     1.38E-09  &   5.15E-09  &    1.03E-08   &   1.80E-07  \\
\enddata           
\end{deluxetable}

\begin{deluxetable}{ccccccc}
\tablewidth{15cm}
\tablecaption{Selected quantities of three mass coordinates of the exploded 35 $\rm M_\odot$ model (set L)}
\tablehead{
\colhead{Mass coordinate} & \colhead{$T_{\rm shock}$} & \colhead{$\rho_{\rm shock}$}
& \colhead{$v_{\rm escape}$} 
& \colhead{$Y_{\rm e}$} & \colhead{Co} & \colhead{Sc} \\
\colhead{$(\rm M_\odot)$} & \colhead{(K)} & \colhead{($\rm g/cm^{3}$)}
& \colhead{(Km/s)} 
& \colhead{} & \colhead{} & \colhead{} \\
}
\startdata
2.01  & 9.53(9) & 1.02(7) & 1.52(4) & 0.499696 & 1.42(-3) & 2.81(-4) \\
2.07  & 7.67(9) & 6.80(6) & 1.38(4) & 0.499697 & 8.61(-3) & 5.84(-6) \\
2.30  & 5.13(9) & 2.89(6) & 1.12(4) & 0.499775 & 4.70(-3) & 9.66(-8) \\
\enddata
\end{deluxetable}

\end{document}